\documentclass{amsart}
\usepackage{hyperref}
\usepackage{graphicx}
\usepackage{curves}

\newcommand{\arxiv}[1]{\href{http://arxiv.org/abs/#1}{arXiv:#1}}
\newcommand*{\mailto}[1]{\href{mailto:#1}{\nolinkurl{#1}}}

\newtheorem{theorem}{Theorem}[section]

\newtheorem{lemma}[theorem]{Lemma}
\newtheorem{corollary}[theorem]{Corollary}
\newtheorem{remark}[theorem]{Remark}
\newtheorem{hypothesis}[theorem]{Hypothesis {\bf H.}\hspace*{-0.6ex}}

\newcommand{\R}{{\mathbb R}}
\newcommand{\N}{{\mathbb N}}
\newcommand{\Z}{{\mathbb Z}}
\newcommand{\C}{{\mathbb C}}
\newcommand{\M}{{\mathbb M}}

\newcommand{\nn}{\nonumber}
\newcommand{\be}{\begin{equation}}
\newcommand{\ee}{\end{equation}}
\newcommand{\bea}{\begin{eqnarray}}
\newcommand{\eea}{\end{eqnarray}}
\newcommand{\ul}{\underline}
\newcommand{\ol}{\overline}
\newcommand{\ti}{\tilde}

\newcommand{\spr}[2]{\langle #1 , #2 \rangle}
\newcommand{\id}{\mathbb{I}}
\newcommand{\I}{\mathrm{i}}
\newcommand{\E}{\mathrm{e}}
\newcommand{\ind}{\mathrm{ind}}
\newcommand{\re}{\mathrm{Re}}
\newcommand{\im}{\mathrm{Im}}

\DeclareMathOperator{\res}{Res}

\def\Xint#1{\mathchoice
   {\XXint\displaystyle\textstyle{#1}}%
   {\XXint\textstyle\scriptstyle{#1}}%
   {\XXint\scriptstyle\scriptscriptstyle{#1}}%
   {\XXint\scriptscriptstyle\scriptscriptstyle{#1}}%
   \!\int}
\def\XXint#1#2#3{{\setbox0=\hbox{$#1{#2#3}{\int}$}
     \vcenter{\hbox{$#2#3$}}\kern-.5\wd0}}
\def\dashint{\Xint-}

\newcommand{\sigI}{\begin{pmatrix} 0 & 1 \\ 1 & 0 \end{pmatrix}}
\newcommand{\rI}{\begin{pmatrix}  1 & 1 \end{pmatrix}}

\newcommand{\ulz}{\underline{z}}
\newcommand{\di}{\mathcal{D}}
\newcommand{\vrc}{\ul{\Xi}_{E_0}}
\newcommand{\hvrc}{\ul{\hat{\Xi}}_{E_0}}
\newcommand{\hmu}{\hat{\mu}}
\newcommand{\hnu}{\hat{\nu}}

\newcommand{\uhnuz}{{\underline{\hat{\nu}}}}

\newcommand{\uhmuz}{{\underline{\hat{\mu}}}}
\newcommand{\dimu}[1]{\di_{\ul{\hat{\mu}}(#1)}}
\newcommand{\dimus}[1]{\di_{\ul{\hat{\mu}}(#1)^*}}
\newcommand{\dimuz}{\di_{\ul{\hat{\mu}}}}
\newcommand{\dimuzs}{\di_{\ul{\hat{\mu}}^*}}
\newcommand{\dinu}[1]{\di_{\ul{\hat{\nu}}(#1)}}
\newcommand{\dinus}[1]{\di_{\ul{\hat{\nu}}(#1)^*}}
\newcommand{\dinuz}{\di_{\ul{\hat{\nu}}}}
\newcommand{\dinuzs}{\di_{\ul{\hat{\nu}}^*}}
\newcommand{\dirho}{\di_{\ul{\rho}}}
\newcommand{\dirhos}{\di_{\ul{\rho}^*}}
\newcommand{\Amap}{\ul{A}_{E_0}}
\newcommand{\amap}{\ul{\alpha}_{E_0}}
\newcommand{\hAmap}{\ul{\hat{A}}_{E_0}}
\newcommand{\hamap}{\ul{\hat{\alpha}}_{E_0}}

\newcommand{\Rg}[1]{R_{2g+1}^{1/2}(#1)}
\newcommand{\Rgo}{R_{2g+1}^{1/2}}

\newcommand{\eps}{\varepsilon}

\newcommand{\sig}{\sigma}
\newcommand{\lam}{\lambda}
\newcommand{\gam}{\gamma}
\newcommand{\om}{\omega}

\numberwithin{equation}{section}

\begin{document}

\title[Asymptotics of Perturbed Finite-Gap KdV Solutions]{Long-Time Asymptotics of Perturbed Finite-Gap Korteweg--de Vries Solutions}

\author[A. Mikikits-Leitner]{Alice Mikikits-Leitner}
\address{Faculty of Mathematics\\ University of Vienna\\
Nordbergstrasse 15\\ 1090 Wien\\ Austria}
\email{\mailto{alice.mikikits-leitner@univie.ac.at}}
\urladdr{\url{http://www.mat.univie.ac.at/~alice/}}

\author[G. Teschl]{Gerald Teschl}
\address{Faculty of Mathematics\\ University of Vienna\\
Nordbergstrasse 15\\ 1090 Wien\\ Austria\\ and International Erwin Schr\"odinger
Institute for Mathematical Physics\\ Boltzmanngasse 9\\ 1090 Wien\\ Austria}
\email{\mailto{Gerald.Teschl@univie.ac.at}}
\urladdr{\url{http://www.mat.univie.ac.at/~gerald/}}

\thanks{Research supported by the Austrian Science Fund (FWF) under Grant No.\ Y330.}
\thanks{J. d'Analyse Math. {\bf 116}, 163--218 (2012)}


\date{\today}

\begin{abstract}
We apply the method of nonlinear steepest descent to compute the long-time asymptotics of 
solutions of the Korteweg--de Vries equation which are decaying perturbations of a quasi-periodic
finite-gap background solution. We compute a nonlinear dispersion relation and show that the $x/t$ plane
splits into $g+1$ soliton regions which are interlaced by $g+1$ oscillatory regions, where $g+1$ is
the number of spectral gaps.

In the soliton regions the solution is asymptotically given by a number of solitons travelling on top
of finite-gap solutions which are in the same isospectral class as the background solution. In the
oscillatory region the solution can be described by a modulated finite-gap solution plus a decaying
dispersive tail. The modulation is given by a phase transition on the isospectral torus and is,
together with the dispersive tail, explicitly characterized in terms of Abelian integrals on the underlying
hyperelliptic curve. 
\end{abstract}

\maketitle

\section{Introduction}

Consider the Korteweg--de Vries (KdV) equation
\be \label{equ:KdVequation}
V_t(x,t)=6V(x,t)V_x(x,t)-V_{xxx}(x,t), \quad (x,t)\in \R\times \R,
\ee
where the subscripts denote differentiation with respect to the corresponding variables.

Following the seminal work of Gardner, Green, Kruskal, and Miura \cite{ggkm}, one can use the inverse scattering transform
to establish existence and uniqueness of (real-valued) classical solutions for the corresponding initial value problem
with rapidly decaying initial conditions. We refer to, for instance, the monograph by Marchenko \cite{mar}. Our concern here
are the long-time asymptotics of such solutions. The classical result is that an arbitrary short-range solution of the
above type will eventually split into a number of solitons travelling to the right plus a decaying radiation part
travelling to the left. The first numerical evidence for such a behaviour was found by Zabusky and Kruskal \cite{zakr}. The first mathematical
results were given by Ablowitz and Newell \cite{an}, Manakov \cite{ma}, and \v{S}abat \cite{sh}. First rigorous results for the
KdV equation were proved by \v{S}abat \cite{sh} and Tanaka \cite{ta2}. Precise asymptotics for the radiation part were first formally
derived by Zakharov and Manakov \cite{zama}, by Ablowitz and Segur \cite{as}, \cite{as2}, by Buslaev \cite{bu} (see also \cite{bb}),
and later on rigorously justified and extended to all orders by Buslaev and Sukhanov \cite{bs}. A detailed rigorous proof
(not requiring any a priori information on the asymptotic form of the solution) was given by Deift and Zhou \cite{dz} based on earlier
work of Manakov \cite{ma} and Its \cite{its1} and is now known as the nonlinear steepest descent
method for oscillatory Riemann--Hilbert problems. For an expository introduction to this method applied to the KdV equation we refer to
\cite{gt}. For further information on the history of this problem we refer to the survey by Deift, Its, and Zhou \cite{diz}.

In this paper we want to look at the case of solutions which are asymptotically close to a quasi-periodic algebro-geometric finite-gap
solution of the KdV equation. In this case the underlying inverse scattering transform was developed only recently by Grunert, Egorova,
and Teschl \cite{et}, \cite{et2}, \cite{egt}. So while the initial value problem for this class of solutions is well understood, nothing was known
about their long-time asymptotics even though the first attempts by Kuznetsov and Mikha\u\i lov \cite{kumi} date back over 35 years ago.
It is the aim of the present paper to fill this gap. In case of the discrete analog, the Toda lattice (see e.g.\ \cite{tjac}),
Kamvissis and Teschl \cite{kt}, \cite{kt2} (with further extensions by Kr\"uger and Teschl \cite{krt3}) have recently extended the
nonlinear steepest descent method for Riemann--Hilbert problem deformations to Riemann surfaces and used this
extension to prove the following result for the Toda lattice:

Let $g$ be the genus of the hyperelliptic curve associated with the unperturbed solution. Then, apart from the
phenomenon of the solitons travelling on the quasi-periodic background, the $(n,t)$-plane contains  $g+2$ areas
where the perturbed solution is close to a finite-gap solution from the same isospectral torus.
In between there are $g+1$ regions where the perturbed solution is asymptotically close to a
modulated lattice which undergoes a continuous phase transition (in the Jacobian variety) and which interpolates
between these isospectral solutions. In the special case of  the free lattice ($g=0$) the
isospectral torus consists of just one point and the known results are recovered.
Both the solutions in the isospectral torus and the phase transition were explicitly
characterized in terms of Abelian integrals on the underlying hyperelliptic curve.

In the present paper we will use this extension for Riemann--Hilbert problems on Riemann surfaces to
prove an analog result for the KdV equation to be formulated in the next section.

\section{Main results}

To set the stage we will choose a quasi-periodic algebro-geometric finite-gap background
solution $V_q(x,t)$ of the KdV equation (cf.\ the next section) plus another solution $V(x,t)$ of
the KdV equation such that
\be \label{decay}
\int_{-\infty}^{+\infty}(1+|x|)^7(|V(x,t)-V_q(x,t)|)dx<\infty
\ee
for all $t\in\R$. We remark that such solutions exist which can be shown by solving
the associated Cauchy problem via the inverse scattering transform \cite{et}.

To fix our background solution $V_q$, let us consider a hyperelliptic Riemann surface $\mathcal{K}_g$ of genus $g\in \N_0$
with real moduli $E_0,E_1,\dots,E_{2g}$. Then we choose a Dirichlet divisor $\dimu{x,t}$ and introduce 
\be
\aligned
\ulz(p,x,t) &= \hvrc - \hAmap(p) + \hamap(\dimu{x,t}) \in \C ^g, \\
\hamap(\dimu{x,t}) &= \hamap(\dimuz) +\frac{x}{2\pi}\ul{U}_0+12\frac{t}{2\pi}\ul{U}_2,
\endaligned
\ee
where $\Amap$ ($\amap$) is Abel's map (for divisors), and $\vrc$, $\ul{U}_0$, and $\ul{U}_2$ are some constants defined
in detail in Section~\ref{secAG} below. Then our background solution is given in terms of Riemann theta functions
(cf.\ \eqref{eq:thetafct}) by
\be
\aligned
V_q(x,t)&=E_0+\sum_{j=1}^g (E_{2j-1}+E_{2j}-2\mu_j(x,t)) \\
&=E_0+\sum_{j=1}^g (E_{2j-1}+E_{2j}-2\lambda_j) -2\partial_x^2\ln \theta \big( \ulz(p_{\infty},x,t)\big),
\endaligned
\ee
where $\lambda_j\in (E_{2j-1},E_{2j})$, $j=1,\dots,g$.

In order to state our main result, we begin by recalling that the perturbed KdV solution $V(x,t)$, $x\in\R$,
for fixed $t\in\R$, is uniquely determined by its scattering data, that is, by the right reflection
coefficient $R_+(\lam,t)$, $\lam\in\sig(H_q)$, and the eigenvalues $\rho_k\in\R\backslash\sig(H_q)$, $k=1,\dots, N$,
together with the corresponding right norming constants $\gam_{+,k}(t)>0$, $k=1,\dots, N$. Here
\be
\sig(H_q) = \bigcup_{j=0}^{g-1}[E_{2j}, E_{2j+1}]\cup[E_{2g},\infty )
\ee
denotes the finite-band spectrum of the underlying background Lax operator
\be
H_q(t) = - \partial_x^2 + V_q(x,t).
\ee
The relation between the energy $\lam$ of the underlying Lax operator $H_q$ and the propagation speed at which
the corresponding parts of the solutions of the KdV equation travel is given by
\be
v(\lam)=\frac{x}{t},
\ee
where
\be
v(\lam) = \lim_{\eps\to 0} \frac{-12\, \re \big( \I \int_{E_0}^{(\lam+\I\eps,+)}\! \omega_{p_{\infty},2}\big)}{ \re \big( \I \int_{E_0}^{(\lam+\I\eps,+)}\! \omega_{p_{\infty},0}\big)},
\ee
and can be regarded as a nonlinear analog of the classical dispersion relation. Here $\omega_{p_{\infty},0}$ and $\omega_{p_{\infty},2}$ are Abelian differentials of the second kind on the underlying Riemann surface defined in \eqref{dsk0} and \eqref{dsk2}.
We will show in Section~\ref{secSPP} that $v$ is a decreasing homeomorphism of $\R$ and we will denote its inverse by $\zeta(v)$.

Furthermore, we define the limiting KdV solution $V_{l,v}(x,t)$ via the relation
\begin{align} \label{limlatsol}
\int_{x}^\infty ( V_{l,v} - V_q )(y,t)dy 
=& -\sum_{\rho_j<\zeta(v)} 4\I \int_{E(\rho_j)}^{\rho_j} \omega_{p_{\infty},0}
+ \frac{1}{\pi} \int_{C(v)} \log(1-|R|^2) \omega_{p_{\infty},0} \nn \\
& {} +2 \partial_x \ln \left( \frac{\theta \big( \ulz(p_{\infty},x,t) + \ul{\delta}(v)\big)}{\theta\big( \ulz(p_{\infty},x,t)\big)} \right),
\end{align}
with
\[
\delta_\ell(v) = -2 \sum_{\rho_j<\zeta(v)} A_{E(\rho_j),\ell}(\rho_j) +
\frac{1}{2\pi\I} \int_{C(v)} \log(1-|R|^2) \zeta_\ell,
\]
where $R=R_+(\lam,t)$ is the associated reflection coefficient and $\zeta_\ell$ is a canonical basis of holomorphic differentials.
Moreover, $C(v)$ is a contour on the Riemann surface obtained by taking the part of the spectrum $\sig(H_q)$ which is to the left of $\zeta(v)$
and lifting it to the Riemann surface (oriented such that the upper sheet lies to its left). Here we have also identified $\rho_j$ with its lift to the
upper sheet and $E(\rho_j)$ denotes the branch point closest to $\rho_j$. If $v=x/t$ we set $V_l(x,t)=V_{l,x/t}(x,t)$.

Then our main result concerning the long-time asymptotics in the soliton region is given by the following theorem:

\begin{theorem} \label{thmMain2}
Assume $V(x,t)$ is a classical solution of the KdV equation \eqref{equ:KdVequation} satisfying
\be 
\int_{-\infty}^{+\infty}(1+|x|^{1+n})(|V(x,t)-V_q(x,t)|)dx<\infty,
\ee
for some integer $n\geq 1$ and abbreviate by $c_k= v(\rho_k)$ the velocity of the $k$'th soliton.
Then the asymptotics in the soliton region, $\{ (x,t) |\, \zeta(x/t) \in\R\backslash\sig(H_q)\}$, are the following:

Let $\eps > 0$ be sufficiently small such that the intervals $[c_k-\eps,c_k+\eps]$, $1\le k \le N$, are disjoint and lie
inside $v(\R\backslash\sig(H_q))$.

If $|\frac{x}{t} - c_k|<\eps$ for some $k$, the solution is asymptotically given by a one-soliton solution on top of the limiting solution:
\be \label{eq:sol1}
\int_x^\infty (V-V_{l,c_k})(y,t) dy = - 2 \frac{\partial}{\partial x} \log\big(c_{l,k}(x,t)\big)+O(t^{-n}),
\ee
as well as
\be \label{eq:sol2}
(V-V_{l,c_k})(x,t) = 2 \frac{\partial^2}{\partial x^2} \log\big(c_{l,k}(x,t)\big)+O(t^{-n}),
\ee
where
\begin{equation}
c_{l,k}(x,t) = 1 + \tilde{\gam}_k \int_x^\infty \psi_{l,c_k}(\rho_k,y,t)^2 dy
\end{equation}
and
\be \label{eq:gamshift}
\aligned
\tilde{\gam}_k & = \gam_k \left( \frac{\theta(\ulz(\rho_k,0,0)+\ul{\delta}(c_k))}{\theta(\ulz(\rho_k,0,0))} \right)^2
\left( \prod_{\rho_j<\zeta(c_k)} \!\!\! \exp\left(2\int_{E_0}^{\rho_k} \om_{\rho_j\, \rho_j^*} \right) \right)\cdot\\
& \quad \cdot  \exp\left( \frac{-1}{\pi\I} \int_{C(c_k)} \!\!\! \log (1-|R|^2) \om_{\rho_k\, p_{\infty}}\right).
\endaligned
\ee
Here $\psi_{l,v}(p,x,t)$ denotes the Baker--Akhiezer function corresponding to the limiting KdV
solution $V_{l,v}(x,t)$ and $\om_{p\, q}$ denotes the Abelian differential of the third kind with poles
at $p$ and $q$.

If $|\frac{x}{t} -c_k| \geq \eps$, for all $k$, the solution is asymptotically close to this limiting solution:
\be
\int_x^\infty (V-V_l)(y,t) dy = O(t^{-n}),
\ee
as well as
\be
V(x,t) = V_l (x,t) + O(t^{-n}).
\ee
\end{theorem}

In particular, we see that the solution splits into a sum of independent solitons where the presence of the other solitons and the radiation part corresponding to the continuous spectrum manifests itself in phase shifts given by \eqref{eq:gamshift}. Moreover, observe that in the periodic case considered here one can have a {\em stationary soliton} (see the discussion in Section~\ref{secSPP}).

The proof will be given at the end of Section~\ref{secSPP}.

\begin{theorem}\label{thmMain3}
Assume $V(x,t)$ is a classical solution of the KdV equation \eqref{equ:KdVequation} satisfying
\eqref{decay} and let $D_j$ be the sector $D_j=\{ (x,t): \zeta(x/t)\in [E_{2j}+\varepsilon,E_{2j+1}-\varepsilon]\}$ for some $\varepsilon >0$. Then the asymptotic is given by
\be
\int_x^{+\infty}(V-V_l)(y,t)dy=4\sqrt{\frac{\I }{\phi ''(z_j)t}}\re \big(\beta (x,t)\big) \Lambda^{\uhnuz}_1(z_j)+O(t^{-\alpha}),
\ee
respectively
\be\label{eqvvl}
(V-V_l)(x,t)= 4\sqrt{\frac{\I }{\phi ''(z_j)t}}\Big[\im \big(\beta (x,t)\big)
-\I\re \big(\beta (x,t)\big)\sum_{k=1}^g\sum_{\ell =1}^g c_{k\ell}(\uhnuz) \zeta_k(z_j)\Big]+O(t^{-\alpha})
\ee
for any $1/2<\alpha<1$ uniformly in $D_j$ as $t\to \infty$.
Here
\be
\phi(p) =
-24 \I \int_{p_0}^p \omega_{p_{\infty},2} - 2\I \frac{x}{t} \int_{p_0}^p \omega_{p_{\infty},0}
\ee
is the phase function,
\be
\frac{\phi ''(z_j)}{\I }=-\frac{12\prod _{k=0,k\neq j}^g(z_j-z_k)}{\I\Rg{z_j}}>0,
\ee
(where $\Rg{z}$ the square root of the underlying Riemann surface $\mathcal{K}_g$ and we identify $z_j$ with its lift to the upper sheet),
\be
\Lambda^\uhnuz_1 (z_j)= \om_{p_{\infty},0}(z_j) - \sum_{k=1}^g\sum_{\ell =1}^g c_{k\ell}(\uhnuz) \alpha_{g-1}(\hat\nu_{\ell}) \zeta_k(z_j),
\ee
with $\om_{p_{\infty},0}$ an Abelian differential of the second kind with a second order pole at $p_{\infty}$ (cf.~eq.~\eqref{dsk0}),
and $\om(p)$ denoting the value of a differential evaluated at $p$ in the chart given by the canonical projection, and
$c_{k\ell}(\uhnuz)$, $\alpha_{g-1}(\hat\nu_{\ell})$ some constants defined in \eqref{defcjlnu}, \eqref{equ:diff2kal}, respectively.
Moreover,
\begin{align}
\beta(x,t) =& \sqrt{\nu} \E^{\I\big( \pi/4-\arg (R(z_j)) +\arg(\Gamma(\I\nu))+2\nu \alpha(z_j)\big)} 
\Big( \frac{\phi ''(z_j)}{\I }\Big)^{-\I \nu}\E^{-t \phi(z_j)} t^{-\I\nu}\cdot \nn \\
&\cdot \frac{\theta \big( \ul{z}(z_j,0,0)\big)}{\theta \big( \ul{z}(z_j,x,t)+\ul{\delta}(x/t)\big)}
\frac{\theta \big( \ul{z}(z_j^*,x,t)+\ul{\delta}(x/t)\big)}{\theta \big( \ul{z}(z_j^*,0,0)\big)}\cdot \nn \\
&\cdot \exp \bigg( -\sum_{\rho_k<\zeta(x/t)}\int_{E(\rho_k)}^{\rho_k}\om_{z_j\,z_j^*}+\frac{1}{2\pi \I}\int_{C(x/t)}\log \Big( \frac{1-|R|^2}{1-|R(z_j)|^2}\Big)\om_{z_j\,z_j^*}\bigg) ,
\end{align}
where $\Gamma (z)$ is the gamma function, $\om_{z_j\,z_j^*}$ an Abelian differential of the third kind defined in \eqref{diffpps},
\be
\nu=-\frac{1}{2\pi }\log \big( 1-|R(z_j)|^2\big)>0,
\ee
and $\alpha (z_j)$ is a constant defined in \eqref{defalphazj}.
\end{theorem}

The proof of this theorem will be given in Section~\ref{secCROSS} of this paper.

Finally, note that if $q(x,t)$ solves the KdV equation, then so does $q(-x,-t)$. Therefore it suffices 
to investigate the case $t\to\infty$.

\section{Algebro-geometric quasi-periodic finite-gap solutions}
\label{secAG}

This section presents some well-known facts on the class of algebro-geometric quasi-periodic finite-gap solutions, that is the class of stationary solutions of the KdV hierarchy, since we want to choose our background solution $V_q$ from that class. We will use the same notation as in \cite{gesztesy2003sea}, where we also refer to for proofs. As a reference for Riemann surfaces in this context we recommend \cite{fk}.

To set the stage let $\mathcal{K}_g$ be the Riemann surface associated with the following function
\begin{equation}\label{eq:r12}
\Rg{z} = \I \prod_{j=0}^{2g} \sqrt{z-E_j},\qquad
E_0 < E_1 < \cdots < E_{2g},
\end{equation}
where $g\in \N_{0}$ and $\{ E_j\}_{j=0}^{2g}$ is a set of real numbers. Here $\sqrt{.}$ denotes the standard root with branch cut along $(0,\infty)$.
We extend $\Rg{z}$ to the branch cuts by setting $\Rg{z} = \lim_{\eps \downarrow 0} \Rg{z + \I \eps}$ for $z \in \C \backslash \Pi$. Hence we have
\be \label{sgnR}
\Rg{z} =\vert \Rg{z} \vert \cdot \left\{
\begin{array}{ll}
(-1)^{g+1} &\mathrm{for} \ z\in (-\infty, E_0),\\
(-1)^{g+j}\I & \mathrm{for}\ z\in (E_{2j},E_{2j+1}), \ j=0,\dots,g-1,\\
(-1)^{g+j} &\mathrm{for}\ z\in (E_{2j+1},E_{2j+2}), \ j=0,\dots,g-1,\\
\I & \mathrm{for} \ z\in (E_{2g},\infty).
\end{array} \right.
\ee
$\mathcal{K}_g$ is a compact, hyperelliptic Riemann surface of genus $g$.

A point on $\mathcal{K}_g$ is denoted by 
$p = (z, \pm \Rg{z}) = (z, \pm)$, $z \in \C$, or $p_{\infty} = (\infty,\infty)$, and
the projection onto $\C \cup \{\infty\}$ by $\pi(p) = z$. 
The points $\{(E_{j}, 0), 0 \leq j \leq 2 g\}\cup \{(\infty,\infty )\} \subseteq \mathcal{K}_g$ are 
called branch points and the sets 
\begin{equation}
\Pi_{\pm} = \{ (z, \pm \Rg{z}) \mid z \in \C\setminus
\bigcup_{j=0}^{g-1}[E_{2j}, E_{2j+1}]\cup[E_{2g},\infty )\} \subset \mathcal{K}_g
\end{equation}
are called upper, lower sheet, respectively.

Next we will introduce the representatives $\{a_j,b_j \}_{j=1}^g$ of a
canonical homology basis for $\mathcal{K}_g$. For $a_j$ we start near $E_{2j-1}$ on
$\Pi_+$, surround $E_{2j}$ thereby changing to $\Pi_-$ and return to our
starting point encircling $E_{2j-1}$ again changing sheets. For $b_j$ we
choose a cycle surrounding $E_0,E_{2j-1}$ counterclockwise (once) on $\Pi_+$.
The cycles are chosen such that their intersection matrix reads
\begin{equation}
a_i \circ a_j= b_i \circ b_j = 0, \qquad a_i \circ b_j = \delta_{i,j},
\qquad 1 \leq i, j \leq g.
\end{equation}







The corresponding canonical basis $\{\zeta_j\}_{j=1}^g$ for the space of
holomorphic differentials can be constructed by
\begin{equation}
\zeta_j = \sum_{k=1}^g c_j(k)  
\frac{\pi^{k-1}d\pi}{\Rgo},
\end{equation}
where the constants $c_j(k)$, $j,k=1,\dots,g$ are given by
\be \label{defcjk}
c_j(k) = C_{jk}^{-1}, \qquad 
C_{jk} = \int_{a_k} \frac{\pi^{j-1}d\pi}{\Rgo} =
2 \int_{E_{2k-1}}^{E_{2k}} \frac{z^{j-1}dz}{\Rg{z}} \in
\R.
\ee
The differentials fulfill
\begin{equation} \label{deftau}
\int_{a_k} \zeta_j = \delta_{k,j}, \qquad \int_{b_k} \zeta_j = \tau_{k,j}, 
\qquad \tau_{k,j} = \tau_{j,k}, \qquad j, k=1,\dots ,g.
\end{equation}
Let us now pick $g$ numbers (the Dirichlet eigenvalues)
\begin{equation}
(\hat{\mu}_j)_{j=1}^g = (\mu_j, \sigma_j)_{j=1}^g
\end{equation}
whose projections lie in the spectral gaps, that is, $\mu_j\in[E_{2j-1},E_{2j}]$, $j=1,\dots,g$. Associated with these numbers is the divisor
\begin{equation}
\dimuz (p)=\left\{ \begin{array}{ll}
1& p=\hat{\mu}_j,\ j=1,\dots,g,\\
0& \mathrm{else}
\end{array}
\right.
\end{equation}
and we can define $g$ numbers $(\hat{\mu}_j(x,t))_{j=1}^g = (\mu_j(x,t), \sigma_j(x,t))_{j=1}^g$ via Jacobi's inversion theorem
by setting
\begin{equation}
\hamap(\dimu{x,t})=  \hamap(\dimuz) +\frac{x}{2\pi}\ul{U}_0+12\frac{t}{2\pi}\ul{U}_2
\end{equation}
such that $\hmu_j(0,0)=\hmu_j$.
Here $\ul{U}_0$ and $\ul{U}_2$ denote the $b$-periods of the Abelian differentials $\omega_{p_{\infty},0}$ and $\omega_{p_{\infty},2}$, respectively, defined below, and $\Amap$ ($\amap$) is Abel's map (for divisors). The hat indicates that we regard it as a (single-valued) map from $\hat{\mathcal{K}}_g$ (the fundamental polygon associated with $\mathcal{K}_g$ by cutting along the $a$ and $b$ cycles) to $\C^g$. 

Next we introduce
\be
\ulz(p,x,t) = \hvrc - \hAmap(p) + \hamap(\dimu{x,t}) \in \C ^g, \\
\ee
where $\hvrc$ is the vector of Riemann constants
\begin{equation}
\hat{\Xi}_{E_0,j} = \frac{j+ \sum_{k=1}^g \tau_{j,k}}{2},\quad j=1,\dots,g.
\end{equation}

By $\theta(\ulz)$ we denote the Riemann theta function associated with $\mathcal{K}_g$ defined by
\begin{equation} \label{eq:thetafct}
\theta(\ulz) = \sum_{\ul{m} \in \Z^g} \exp 2 \pi \I \left( \spr{\ul{m}}{\ulz} +
\frac{\spr{\ul{m}}{\ul{\tau} \, \ul{m}}}{2}\right) ,\qquad \ulz \in \C^g.
\end{equation}
Note that the function $\theta(\ulz(p,x,t))$ has precisely $g$ zeros
$\hmu_j(x,t)$. This follows from Riemann's vanishing theorem (cf.~\cite[Theorem~A.13]{tjac}).

Introduce the time-dependent Baker--Akhiezer function
\begin{equation}\label{defpsiq}
\psi_q(p,x,t) = \frac{\theta \big(\ulz(p,x,t)\big)}{\theta\big(\ulz(p_{\infty},x,t)\big)}\frac{\theta \big(\ulz(p_{\infty},0,0)\big)}{\theta \big(\ulz(p,0,0)\big)}
\exp \Big( -\I x \int_{E_0}^p \omega_{p_{\infty},0} -12\I t\int_{E_0}^p \omega_{p_{\infty},2} 
\Big).
\end{equation}
Here $\omega_{p_{\infty},0}$ and $\omega_{p_{\infty},2}$ are normalized Abelian differentials of the second kind with a single pole at $p_{\infty}$ and principal part $w^{-2}dw$ and $w^{-4}dw$ in the chart $w(p) = \pm \I z^{-1/2}$ for $p=(z,\pm)$, respectively. The Abelian differentials are normalized to have vanishing $a_j$ periods and have the following expressions
\begin{equation}\label{dsk0}
\omega_{p_{\infty},0}=\frac{1}{2\I}\frac{\prod_{j=1}^g(\pi-\lambda_j)}{\Rgo}d\pi,
\end{equation}
with $\lambda_j\in (E_{2j-1},E_{2j})$, $j=1,\dots,g$, and
\begin{equation}\label{dsk2}
\omega_{p_{\infty},2}=\frac{1}{2\I}\frac{\prod_{j=0}^g(\pi-\tilde{\lambda}_j)}{\Rgo}d\pi,
\end{equation}
where $\tilde{\lambda}_j$, $j=0,\dots,g$, have to be chosen such that they fulfill $\sum_{j=0}^{g}\tilde{\lambda}_j=\frac{1}{2}\sum_{j=0}^{2g}E_j$.
We also remark
\be\label{psipps}
\psi_q(p,x,t) \psi_q(p^*,x,t) = \prod_{j=1}^g \frac{z-\mu_j(x,t)}{z-\mu_j}, \qquad p=(z,\pm).
\ee
Then our background KdV solution is given by
\begin{equation}
V_q(x,t)=E_0+\sum_{j=1}^g (E_{2j-1}+E_{2j}-2\lambda_j) -2\partial_x^2\ln \theta \big( \ulz(p_{\infty},x,t)\big).
\end{equation}

The Abelian differentials of the third kind $\omega_{q_1\,q_2}$, with simple poles at $q_1$ and $q_2$, corresponding residues $+1$ and $-1$,
vanishing $a$-periods, and holomorphic on $\mathcal{K}_g\setminus \{ q_1,q_2\}$, are explicitly given by (\cite[Appendix~B]{gesztesy2003sea})
\begin{align}
\omega_{p_1\,p_2}&= 
\Big( \frac{\Rgo +\Rg{p_1}}{2\big(\pi-\pi(p_1)\big)}-\frac{\Rgo +\Rg{p_2}}{2\big(\pi-\pi(p_2)\big)}+P_{p_1\, p_2}(z)\Big)\frac{d\pi}{\Rgo},\\
\omega_{p_1\,p_{\infty}}&= 
\Big(\frac{\Rgo+\Rg{p_1}}{2\big(\pi-\pi(p_1)\big)}+P_{p_1\, p_{\infty}} (z)\Big) \frac{d\pi}{\Rgo},
\end{align}
where $p_1, p_2 \in \mathcal{K}_g\setminus \{ p_{\infty}\}$ and $P_{p_1\, p_2}(z)$ and $P_{p_1\, p_{\infty}}(z)$ are polynomials of degree $g-1$ which have to be determined from the normalization $\int_{a_\ell}\omega_{p_1\,p_2}=0$ and $\int_{a_\ell}\omega_{p_1\,p_{\infty}}=0$, respectively.
In particular,
\be \label{diffpps}
\om_{p p^*} = \Big(\frac{\Rg{p}}{\pi - \pi(p)} + P_{p p^*}(\pi) \Big)
\frac{d\pi}{\Rgo}.
\ee
We will also need the Blaschke factor
\be
B(p,\rho)= \exp\Big(\int_{E_0}^p \om_{\rho\, \rho^*}\Big) =
\exp\Big(\int_{E(\rho)}^\rho \om_{p\, p^*}\Big), \quad \pi(\rho)\in\R,
\ee
where $E(\rho)$ is $E_0$ if $\rho<E_0$, either $E_{2j-1}$ or $E_{2j}$ if
$\rho\in(E_{2j-1},E_{2j})$, $1\le j \le g$.
It is a multivalued function with a simple zero at $\rho$ and simple pole at $\rho^*$
satisfying $|B(p,\rho)|=1$, $p\in\partial\Pi_+$. It is real-valued for $\pi(p)\in(-\infty,E_0)$ and
satisfies
\be\label{eq:propblaschke}
B(E_0,\rho)=1 \quad\mbox{and}\quad
B(p^*,\rho) = B(p,\rho^*) = B(p,\rho)^{-1}
\ee
(see e.g., \cite{tag}).

The Baker--Akhiezer function is a meromorphic function on $\mathcal{K}_g\setminus \{p_{\infty}\}$
with an essential singularity at $p_{\infty}$. The two branches are denoted by
\begin{equation}
\psi_{q,\pm}(z,x,t) = \psi_q(p,x,t), \qquad p=(z,\pm),
\end{equation}
and it satisfies
\begin{align}\nn
H_q(t) \psi_q(p,x,t) &= \pi(p) \psi_q(p,x,t),\\
\frac{d}{dt} \psi_q(p,x,t) &= P_{q,2}(t) \psi_q(p,x,t).
\end{align}
Here 
\begin{align}\nn
H_q(t) &= \partial_x^2+V_q(.,t),\\
P_{q,2}(t) &= -4\partial_x^3+6V_q(.,t)\partial_x+3V_{q,x}(.,t),
\end{align}
are the operators from the Lax pair for the KdV equation, that is,
\be
\frac{d}{dt} H_q(t) = H_q(t) P_{q,2}(t) - P_{q,2}(t) H_q(t).
\ee

It is well known that the spectrum of $H_q(t)$ is time independent and
consists of $g+1$ bands
\begin{equation}
\sig(H_q(t)) = \bigcup_{j=0}^{g-1} [E_{2j},E_{2j+1}]\cup [E_{2g},\infty ).
\end{equation}
For further information and proofs we refer to \cite{gesztesy2003sea}.

\section{The Inverse scattering transform and the Riemann--Hilbert problem}
\label{secISTRH}

In this section we recall some basic facts from the inverse scattering transform for our setting. For further background
and proofs we refer to \cite{bet}, \cite{egt}, and \cite{et} (see also \cite{mikikits2009trace}).

Let $\psi_{q,\pm}(z,x,t)$ be the branches of the Baker--Akhiezer function defined in the previous section. Let $\psi_\pm(z,x,t)$ be the Jost functions for the perturbed problem 
\be
\big( -\partial_x^2+V(x,t)\big) \psi_\pm(z,x,t)=z\psi_\pm(z,x,t),
\ee
defined by the asymptotic normalization
\begin{equation}
\lim_{x \rightarrow \pm \infty}
\E^{\mp \I xk(z)} \big( \psi_{\pm}(z,x,t) - \psi_{q,\pm}(z,x,t) \big) = 0,
\end{equation}
where $k(z)$ denotes the quasimomentum map
\begin{equation}
k(z)=-\int_{E_0}^{p}\omega_{p_{\infty},0}, \qquad p=(z,+).
\end{equation}
The asymptotics of the two projections of the Jost function are (cf.~\cite[Theorem~2.3]{mikikits2009trace})
\begin{equation} \label{asymJost}
\psi_\pm(z,x,t) = \psi_{q,\pm}(z,x,t)\Big(1 \mp\int_{x}^{\pm\infty} ( V-V_q)(y,t)dy \frac{1}{2\I \sqrt{z}} + o(1/\sqrt{z}) \Big), 
\end{equation}
as $z \to \infty$. We will assume that the poles of the Baker--Akhiezer function $\mu_k$ are all different from the
eigenvalues $\rho_j$ without loss of generality (otherwise just shift the base point $(x_0,t_0)=(0,0)$).

One has the scattering relations
\be \label{relscat}
T(z) \psi_\mp(z,x,t) =  \ol{\psi_\pm(z,x,t)} +
R_\pm(z) \psi_\pm(z,x,t),  \qquad z \in\sigma(H_q),
\ee
where $T(z)$, $R_\pm(z)$ are the transmission respectively reflection coefficients.
Here $\psi_\pm(z,x,t)$ is defined such that 
$\psi_\pm(z,x,t)= \lim_{\eps\downarrow 0}\psi_\pm(z + \I\eps,x,t)$,
$z\in\sigma(H_q)$. If we take the limit from the other side we
have $\ol{\psi_\pm(z,x,t)}= \lim_{\eps\downarrow 0}\psi_\pm(z - \I\eps,x,t)$.

The transmission and reflection coefficients have the following well-known properties:

\begin{lemma}
The transmission coefficient $T(z)$ has a meromorphic extension to
$\C\backslash\sig(H_q)$ with simple poles at the eigenvalues $\rho_j$.
The residues of $T(z)$ are given by
\begin{equation} \label{eq:resT}
\res_{\rho_j} T(z) = \frac{2\Rg{\rho_j}}{\prod_{k=1}^g (\rho_j-\mu_k)} \frac{\gam_{\pm,j}}{c_j^{\pm 1}},
\end{equation}
where
\be
\gam_{\pm,j}^{-1} = \int_{-\infty}^{\infty} |\psi_\pm(\rho_j,y,t)|^2dy
\ee
are referred to as norming constants and $\psi_- (\rho_j,x,t) = c_j \psi_+(\rho_j,x,t)$.

Moreover,
\be \label{reltrpm} 
T(z) \ol{R_+(z)} + \ol{T(z)} R_-(z)=0, \qquad |T(z)|^2 + |R_\pm(z)|^2=1.
\ee
\end{lemma}

In particular, one reflection coefficient, say $R(z)=R_+(z)$, and one set of
norming constants, say $\gam_j= \gam_{+,j}$, will be sufficient for us.

We will define a sectionally meromorphic vector on the Riemann surface $\mathcal{K}_g$ as follows:
\be\label{defm}
m(p,x,t)= \left\{\begin{array}{c@{\quad}l}
\begin{pmatrix} T(z) \frac{\psi_-(z,x,t)}{\psi_{q,-}(z,x,t)}  & \frac{\psi_+(z,x,t)}{\psi_{q,+}(z,x,t)} \end{pmatrix},
& p=(z,+)\\
\begin{pmatrix} \frac{\psi_+(z,x,t)}{\psi_{q,+}(z,x,t)} & T(z) \frac{\psi_-(z,x,t)}{\psi_{q,-}(z,x,t)} \end{pmatrix}, 
& p=(z,-)
\end{array}\right..
\ee
We are interested in the jump condition of $m(p,x,t)$ on $\Sigma$, the boundary of $\Pi_\pm$ (oriented counterclockwise when viewed from top sheet $\Pi_+$). It consists of two copies $\Sigma_\pm$ of $\sigma(H_q)$ which correspond to non-tangential limits from $p=(z,+)$ with $\pm\im(z)>0$, respectively to non-tangential limits from $p=(z,-)$ with $\mp\im(z)>0$.

To formulate our jump condition we use the following convention: When representing functions on $\Sigma$, the lower subscript denotes the non-tangential limit from $\Pi_+$ or $\Pi_-$, respectively,
\begin{equation}
m_\pm(p_0) = \lim_{ \Pi_\pm \ni p\to p_0} m(p), \qquad p_0\in\Sigma.
\end{equation}
Using the notation above implicitly assumes that these limits exist in the sense that $m(p)$ extends to a continuous function on the boundary away from the band edges.

Moreover, we will also use symmetries with respect to the sheet exchange map
\begin{equation}
p^*= \begin{cases}
(z,\mp) & \text{ for } p=(z,\pm),\\
p_\infty & \text{ for } p=p_\infty,
\end{cases}
\end{equation}
and complex conjugation
\begin{equation}
\ol{p} = \begin{cases}
(\ol{z},\pm) & \text{ for } p=(z,\pm)\not\in \Sigma,\\
(z,\mp) & \text{ for } p=(z,\pm)\in \Sigma,\\
p_\infty & \text{ for } p= p_\infty.
\end{cases}
\end{equation}
In particular, we have $\ol{p}=p^*$ for $p\in\Sigma$.

Note that we have $\ti{m}_\pm(p)=m_\mp(p^*)$ for $\ti{m}(p)= m(p^*)$ (since $*$ reverses the orientation of $\Sigma$) and $\ti{m}_\pm(p)= \ol{m_\pm(p^*)}$ for $\ti{m}(p)=\ol{m(\ol{p})}$.

Note that we have the following asymptotic behavior for $m(p,x,t)$ near $p_{\infty}$:
\begin{equation} \label{m2infp}
m(p)=\begin{pmatrix} 1 & 1 \end{pmatrix}
-\frac{1}{2\I \sqrt{z}}\int_{x}^{\infty} ( V-V_q)(y,t)dy \begin{pmatrix} -1 & 1 \end{pmatrix} +o\big( \frac{1}{\sqrt{z}}\big), \quad p=(z,\pm )
\end{equation}
for $p$ near $p_{\infty }$. Here we made use of \eqref{asymJost} and
\begin{equation}
T(z) = 1 + \frac{1}{2\I \sqrt{z}} \int_{-\infty}^{\infty} (V-V_q)(y,t) dy + o\big(\frac{1}{\sqrt{z}}\big)
\end{equation}
(cf.~\cite[Corollary~3.7]{mikikits2009trace}).

We are now ready to derive the main vector Riemann--Hilbert problem:

\begin{theorem}[Vector Riemann--Hilbert problem]\label{thm:vecrhp}
Let $\mathcal{S}_+(H(0))=\{ R(\lam),\; \lam\in\sig(H_q); \: (\rho_j, \gam_j), \: 1\le j \le N \}$
the right scattering data of the operator $H(0)$. Then $m(p)=m(p,x,t)$ defined in \eqref{defm}
is meromorphic away from $\Sigma$ and satisfies:
\begin{enumerate}
\item The jump condition
\begin{equation} \label{eq:jumpcond}
m_+(p)=m_-(p) J(p), \qquad
J(p)= \begin{pmatrix}
1 -|R(p)|^2  &- \ol{R(p) \Theta(p,x,t)}  \E^{-t \phi(p)}\\
R(p)  \Theta(p,x,t) \E^{t \phi(p)} & 1
\end{pmatrix},
\end{equation}
for $p\in\Sigma$,
\item
the divisor
\begin{equation} \label{eq:divcond}
(m_1) \ge -\dimus{x,t} - \dirho, \qquad (m_2) \ge -\dimu{x,t} - \dirhos
\end{equation}
and pole conditions
\begin{equation} \label{eq:polecond}
\aligned
& \Big( m_1(p) + \frac{-2\Rg{\rho_j}}{\prod_{k=1}^g (\rho_j-\mu_k)}
\frac{\gam_j}{\pi(p)-\rho_j} \frac{\psi_q(p,x,t)}{\psi_q(p^*,x,t)} m_2(p) \Big) \ge - \dimus{x,t},
\mbox{ near $\rho_j$},\\
& \Big( \frac{-2\Rg{\rho_j}}{\prod_{k=1}^g (\rho_j-\mu_k)}
\frac{\gam_j}{\pi(p)-\rho_j} \frac{\psi_q(p^*,x,t)}{\psi_q(p,x,t)}  m_1(p) + m_2(p) \Big) \ge - \dimu{x,t},
\mbox{ near $\rho_j^*$},
\endaligned
\end{equation}
\item
the symmetry condition
\begin{equation} \label{eq:symcond}
m(p^*) = m(p) \sigI 
\end{equation}
\item
and the normalization
\begin{equation} \label{eq:normcond}
m(p_{\infty})= \begin{pmatrix}1 & 1 \end{pmatrix}.
\end{equation}
\end{enumerate}
Here $(f)$ denotes the divisor of $f$ and
\begin{equation}
\dirho= \sum_j \di_{\rho_j}, \qquad \dirhos= \sum_j \di_{\rho_j^*}.
\end{equation}
denotes the divisor corresponding to the points $\rho_j\equiv(\rho_j,+)\in\mathcal{K}_g$. The phase $\phi$ is given by
\begin{equation} \label{defsp}
\phi(p,\frac{x}{t}) =
-24 \I \int_{p_0}^p \omega_{p_{\infty},2} - 2\I \frac{x}{t} \int_{p_0}^p \omega_{p_{\infty},0}
\in \I \R \quad \text{for }p\in \Sigma.
\end{equation}
Moreover, we have set
\begin{equation} \label{defTheta}
\Theta(p,x,t) = \frac{\theta(\ulz(p,x,t))}{\theta(\ulz(p,0,0))}
\frac{\theta(\ulz(p^*,0,0))}{\theta(\ulz(p^*,x,t))}
\end{equation}
such that
\[
\frac{\psi_q(p,x,t)}{\psi_q(p^*,x,t)} = \Theta(p,x,t) \E^{t\phi(p)}.
\]
\end{theorem}

Here we have extended our definition of $R$ to $\Sigma$ such that it is equal to $R(z)$ on $\Sigma_+$ and equal to $\ol{R(z)}$ on $\Sigma_-$.
In particular, the condition on $\Sigma_+$ is just the complex conjugate of the one on $\Sigma_-$ since we have $R(p^*)= \ol{R(p)}$
and $m_\pm(p^*,x,t)= \ol{m_\pm(p,x,t)}$ for $p\in\Sigma$.

\begin{proof}
The jump condition follows by using \eqref{relscat} and \eqref{reltrpm}. By Riemann's vanishing theorem (cf.~\cite[Theorem~A.13]{tjac}) the Baker--Akhiezer function $\psi_q$ has simple zeros at $\hat{\mu}_j(x,t)$ and simple poles at $\hat{\mu}_j$, $j=1,\dots,g$. Moreover, the transmission coefficient $T(z)$ has simple poles at the eigenvalues $\rho_j$, $j=1,\dots,N$. Thus the divisor conditions \eqref{eq:divcond} are indeed fulfilled. The pole conditions follow from the fact that the transmission coefficient $T(z)$ is meromorphic in $\C\setminus \sig(H_q)$ with simple poles at $\rho_j$ and its residues are given by \eqref{eq:resT}. The symmetry condition \eqref{eq:symcond} obviously holds by the definition of the function $m(p)$. The normalization \eqref{eq:normcond} is immediately clear from \eqref{m2infp}.
\end{proof}

We note that the symmetry condition is in fact crucial to guarantee that the solution of this vector Riemann--Hilbert problem is unique. 

\begin{theorem}
The vector $m(p)$ defined in \eqref{defm} is the only solution of the vector Riemann--Hilbert problem
\eqref{eq:jumpcond}--\eqref{eq:normcond}.
\end{theorem}

\begin{proof}
The argument is similar to \cite[Thm.~B.1]{kt2}.
It suffices to show that the corresponding vanishing Riemann--Hilbert problem, where the normalization
condition \eqref{eq:normcond} is replaced by $m(p_{\infty})= \begin{pmatrix}0 & 0 \end{pmatrix}$, has only the trivial solution.

Let $\ti{m}$ be some solution of the vanishing Riemann--Hilbert problem.
We want to apply Cauchy's integral theorem to $\ti{m}(p) \ti{m}^\dag(\ol{p}^*)$.
To handle the poles of $\ti{m}$ we will multiply it by a meromorphic differential $d\Omega$
which has zeros at $\ul{\mu}$ and $\ul{\mu}^*$ and a simple pole
at $p_{\infty}$ such that finally the differential $\ti{m}(p) \ti{m}^\dag(\ol{p}^*) d\Omega(p)$
is holomorphic away from the contour. Here $\ti{m}^\dag$ denotes the adjoint (transpose and complex conjugate)
vector of $\ti{m}$.

More precisely, let 
\be
d\Omega= \frac{\prod_{j=1}^g (\pi -\mu_j)}{-R_{2g+1}^{1/2}} d\pi
\ee
and note that $-\big( \prod_j(z-\mu_j) \big)R^{-1/2}_{2g+1}(z)$ is a Herglotz
function. That is, it has positive imaginary part in the upper half-plane (and it is
purely imaginary on $\sig(H_q)$). Hence $ \ti{m} (p) \ol{\ti{m}^T(p)}d\Omega(p)$ will be positive
on $\Sigma$.

Next, consider the integral
\be\label{eq:intD}
0= \int_D \ti{m}(p)  \ti{m}^\dag(\ol{p}^*) d\Omega(p),
\ee
where $D$ is a $\ol{*}$-invariant contour consisting of two loops on the upper and on the lower sheet encircling none
of the poles $\rho_j$, $\rho_j^*$. We first deform $D$ to a $\ol{*}$-invariant contour consisting of several parts:
Two pieces $D_\pm$ wrapping around the $\pm$ side of $\Sigma$ plus a number of small circles $D_{+,j}$, $D_{-,j}$
around the poles $\rho_j$, $\rho_j^*$, respectively. Then the contribution from $\Sigma$ is given by
\begin{align} \nn
\int_\Sigma \ti{m}(p) \ti{m}^\dag(\ol{p}^*) d\Omega(p)& = \int_\Sigma \big( \ti{m}_+(p) \ti{m}_-^\dag(\ol{p}^*) + \ti{m}_-(p) \ti{m}_+^\dag(\ol{p}^*)\big) d\Omega(p) \\ 
&= \int_\Sigma \ti{m}_-(p) (J(p) +J^\dag(\ol{p}^*)) \ti{m}_-^\dag(\ol{p}^*) d\Omega(p)\geq 0
\end{align}
and the contribution from the poles is given by
\begin{align} \nn
& \int_{\cup_{j=1}^N (D_{+,j} \cup D_{-,j})} \ti{m}(p) \ti{m}^\dag(\ol{p}^*) d\Omega (p)\\\nn
&\quad = \sum_{j=1}^N \left( \res_{\rho_j} \ti{m}(p) \ti{m}^\dag(\ol{p}^*)d\Omega(p) + \res_{\rho_j^*} \ti{m}(p) \ti{m}^\dag(\ol{p}^*) d\Omega(p)\right)\\
&\quad = 2 \sum_{j=1}^N\res_{\rho_j} \ti{m}(p) \ti{m}^\dag(\ol{p}^*) d\Omega(p).
\end{align}
To compute the residues we use the pole conditions \eqref{eq:polecond} which imply (using \eqref{psipps})
\[
\res_{\rho_j} \ti{m}(p) \ti{m}^\dag(\ol{p}^*) d\Omega(p) = \frac{2 \gam_j}{\prod_{k=1}^g(\rho_j-\mu_k^0)^2} \psi_q(\rho_j)^2 m_2(\rho_j)^2 \ge 0.
\]
In particular, both contributions to the integral \eqref{eq:intD} are non-negative and thus both must vanish.
It follows from the that $\ti{m} = 0$ vanishes along $\Sigma$ and consequently $\ti{m}(p)=0$ as desired.
\end{proof}

We will also need another asymptotic relation
\begin{equation} \label{equ:expm1m2}
m_1\cdot m_2=1+( V-V_q)(x,t)\frac{1}{2z}+o(z^{-1}).
\end{equation}
which is immediate from the following well-known result.

\begin{lemma} \label{lem:expTpsipsiq}
We have
\be
T(z)\frac{\psi_-(z,x,t)}{\psi_{q,-}(z,x,t)}\frac{\psi_+(z,x,t)}{\psi_{q,+}(z,x,t)}=1+\frac{1}{2}(V-V_q)(x,t)\frac{1}{z}+o(z^{-1}).
\ee
\end{lemma}

\begin{proof}
We will use the following representation of the Jost solutions
\be\label{Jost1}
\psi_\pm(z,x,t)=\psi_{q,\pm}(z,x,t)\exp\Big(\mp\int_x^{\pm\infty}
\big( m_\pm(z,y,t) - m_{q,\pm}(z,y,t)\big) dy \Big),
\ee
where 
\[
m_{\pm}(z,x,t)= \pm\frac{\psi_{\pm}'(z,x,t)}{\psi_{\pm}(z,x,t)}, \quad m_{q,\pm}(z,x,t)= \pm\frac{\psi_{q,\pm}'(z,x,t)}{\psi_{q,\pm}(z,x,t)}
\]
are the Weyl--Titchmarsh functions. Here the prime denotes differentiation with respect to $x$. Using the expansion of the Weyl $m$-functions (cf.~\cite[Lemma~6.1]{mikikits2009trace}) and the one for $\log T(z)$ (cf.~ \cite[Theorem~6.2]{mikikits2009trace}) for $z\to \infty$ proves the claim.
\end{proof}

For our further analysis it will be convenient to rewrite the pole conditions as jump conditions following the idea of Deift, Kamvissis, Kriecherbauer, and Zhou \cite{dkkz}. For that purpose we choose $\eps$ so small that the discs $|\pi(p)-\rho_j|<\eps$ are inside the upper sheet $\Pi_+$ and do not intersect with the spectral bands. Then redefine $m(p)$ in a neighborhood of $\rho_j$ respectively $\rho_j^*$ in the following way:
\begin{equation} \label{eq:redefm}
m(p) = \begin{cases}
m(p) \begin{pmatrix} 1 & 0 \\ 
\frac{\gam_j(p,x,t)}{\pi(p)-\rho_j} & 1 \end{pmatrix},  &
\begin{smallmatrix}|\pi(p)-\rho_j|<\eps\\ p\in\Pi_+\end{smallmatrix},\\
m(p) \begin{pmatrix} 1 & 
\frac{\gam_j(p^*,x,t)}{\pi(p)-\rho_j} \\
0 & 1 \end{pmatrix},  &
\begin{smallmatrix}|\pi(p)-\rho_j|<\eps\\ p\in\Pi_-\end{smallmatrix},\\
m(p), & \text{else},\end{cases}
\end{equation}
where $\gam_j(p,x,t)$ is a function which is analytic in $0<|\pi(p)-\rho_j|<\eps$, $p\in\Pi_+$ and satisfies
\begin{equation*}
\lim_{p\to \rho_j}\gam_j(p,x,t)\frac{\psi_q(p^*,x,t)}{\psi_q(p,x,t)}=
\frac{2\Rg{\rho_j}}{\prod_{k=1}^g (\rho_j-\mu_k)} \gam_j.
\end{equation*}
For example, we can choose 
\begin{equation*}
\gam_j(p,x,t)=\frac{-2\Rg{\rho_j}}{\prod_{k=1}^g (\rho_j-\mu_k)}\frac{\psi_q(p,x,t)}{\psi_q(p^*,x,t)} \gam_j
\end{equation*}
or
\begin{equation*}
\gam_j(p,x,t)=\frac{-2\Rg{\rho_j}}{\prod_{k=1}^g (\pi(p)-\mu_k)}\frac{\psi_q(p,x,t)}{\psi_q(p^*,x,t)} \gam_j.
\end{equation*}

\begin{lemma}\label{lem:pctoj}
Suppose $m(p)$ is redefined as in \eqref{eq:redefm}. Then $m(p)$ is meromorphic away from
$\Sigma$ and satisfies \eqref{eq:jumpcond}, \eqref{eq:symcond}, \eqref{eq:normcond},
the divisor conditions change according to
\begin{equation}
(m_1) \ge -\dimus{x,t}, \qquad (m_2) \ge -\dimu{x,t}
\end{equation}
and the pole conditions are replaced by the jump conditions
\begin{equation} \label{eq:jumpcond2}
\aligned
m_+(p) &= m_-(p) \begin{pmatrix} 1 & 0 \\
\frac{\gam_j(p,x,t)}{\pi(p)-\rho_j} & 1 \end{pmatrix},\quad p\in\Sigma_\eps(\rho_j),\\
m_+(p) &= m_-(p) \begin{pmatrix} 1 & -\frac{\gam_j(p^*,x,t)}{\pi(p)-\rho_j} \\
0 & 1 \end{pmatrix},\quad
p\in\Sigma_\eps(\rho_j^*),
\endaligned
\end{equation}
where 
\begin{equation}
\Sigma_\eps(p) = \{ q \in \Pi_\pm\,:\, |\pi(q)-z| =\eps \}, \qquad p=(z,\pm),
\end{equation}
is a small circle around $p$ on the same sheet as $p$. It is oriented counterclockwise on the upper sheet and clockwise on the lower sheet.
\end{lemma}

\begin{proof}
Everything except for the pole conditions follows as in the proof of Theorem~\ref{thm:vecrhp}. That the pole conditions \eqref{eq:polecond} are indeed replaced by the jump conditions \eqref{eq:jumpcond2} as $m(p)$ is redefined as in \eqref{eq:redefm} can be shown by a straightforward calculation.
\end{proof}

The next thing we will do will be to deduce the one-soliton solution of our Riemann--Hilbert problem, i.e., the solution in the case where only one eigenvalue $\rho$ corresponding to one bound state is present and the reflection coefficient $R(p)$ vanishes identically on $\mathcal{K}_g$.

\begin{lemma}[One-soliton solution] \label{lem:singlesoliton}
Suppose there is only one eigenvalue and a vanishing reflection coefficient, that is,
$\mathcal{S}_+(H(t))=\{ R(p)\equiv 0,\; p\in\Sigma; \: (\rho, \gam) \}$.
Let
\begin{equation}
c_{q,\gam}(\rho,x,t) = 1 + \gam W_{(x,t)}(\dot{\psi}_q(\rho,x,t),\psi_q(\rho,x,t))
= 1 + \gam \int_x^\infty \psi_q(\rho,y,t)^2 dy
\end{equation}
and
\begin{equation}
\psi_{q,\gam}(p,x,t) = \psi_q(p,x,t) +\frac{\gam}{z-\rho} \frac{\psi_q(\rho,x,t)
W_{(x,t)}(\psi_q(\rho,x,t),\psi_q(p,x,t))}{c_{q,\gam}(\rho,x,t)}.
\end{equation}
Here the dot denotes a derivate with respect to $\rho$ and $W_{(x,t)}(f,g)=(f(x)g'(x)-f'(x)g(x))$ is the usual Wronski determinant, where the prime denotes the derivative with respect to $x$.

Then the unique solution of the Riemann--Hilbert problem \eqref{eq:jumpcond}--\eqref{eq:normcond}
is given by
\be\label{eq:oss}
m_0(p) = \begin{pmatrix} f(p^*,x,t) & f(p,x,t) \end{pmatrix}, \qquad
\nn f(p,x,t) = \frac{\psi_{q,\gam}(p,x,t)}{\psi_q(p,x,t)}.
\ee
In particular
\be \label{equ:vminvqsol}
\int_x^{\infty}(V-V_q)(y,t)dy = -2 \frac{\partial}{\partial x} \log\big(c_{q,\gam}(\rho,x,t) \big),
\ee
or
\be \label{equ:vminvqsol2}
(V-V_q)(x,t) = 2 \frac{\partial^2}{\partial x^2} \log\big(c_{q,\gam}(\rho,x,t) \big).
\ee
\end{lemma}

\begin{proof}
Since we assume the reflection coefficient vanishes, the jump along $\Sigma$ disappears. Moreover, since the symmetry condition \eqref{eq:symcond} has to be satisfied it follows that the solution of the Riemann--Hilbert problem \eqref{eq:jumpcond}--\eqref{eq:normcond} has to be of the form $m_0(p) = \begin{pmatrix} f(p^*,x,t) & f(p,x,t) \end{pmatrix}$. The divisor conditions \eqref{eq:divcond} follow from the fact that the Baker--Akhiezer function $\psi_q$ has simple zeros at $\hat{\mu}_j(x,t)$ and simple poles at $\hat{\mu}_j$, $j=1,\dots,g$ and by construction of $\psi_{q,\gamma}$. It is obvious that the normalization condition \eqref{eq:normcond} holds. Thus it is only left to check the pole conditions \eqref{eq:polecond}. For this purpose we compute
\begin{align*}
\lim_{p\to \rho} (z-\rho) f(p^*) &= \frac{\gam(\rho,x,t)}{c_{q,\gam}(\rho,,x,t)}
W_{(x,t)}(\psi_q(\rho,x,t),\psi_q(\rho^*,x,t))\\
&= -\frac{\gam(\rho,x,t)}{c_{q,\gam}(\rho,,x,t)}\frac{2\Rg{\rho}}{\prod_{k=1}^g (\rho-\mu_k)},
\end{align*}
where we defined
\[
\gam(p,x,t)= \gam \frac{\psi_q(p,x,t)}{\psi_q(p^*,x,t)} = \gam \Theta(p,x,t) \E^{t \phi(p)},
\]
and we used (cf.~\cite[Equ.~(1.87)]{gesztesy2003sea})
\begin{equation*}
W(\psi_{q,\mp}(z),\psi_{q,\pm}(z))=\pm\frac{2\Rg z}{\prod_{k=1}^g(z-\mu_k)}.
\end{equation*}
Moreover,
\begin{align*}
\lim_{p\to \rho}f(p) = 1&+\frac{\gam}{c_{q,\gam}(\rho,x,t)}\lim_{p\to \rho}\frac{W_{(x,t)}(\psi_q(\rho,x,t),\psi_q(p,x,t))}{z-\rho}\\
= 1&+\frac{\gam}{c_{q,\gam}(\rho,x,t)} \big[ \psi_q(\rho,x,t)\lim_{p\to \rho}\frac{\psi_q'(p,x,t)-\psi_q'(\rho,x,t)}{z-\rho}-\\ &- \psi_q'(\rho,x,t)\lim_{p\to \rho}\frac{\psi_q(p,x,t)-\psi_q(\rho,x,t)}{z-\rho}\big] \\
= 1&+ \frac{\gam}{c_{q,\gam}(\rho,x,t)}W_{(x,t)}(\psi_q(\rho,x,t),\dot{\psi}_q(\rho,x,t))= \frac{1}{c_{q,\gam}(\rho,x,t)}.
\end{align*}
Hence we see that the pole conditions \eqref{eq:polecond} are satisfied.

The formula \eqref{equ:vminvqsol} follows after expanding around $p=p_{\infty}$, that is, 
\begin{align*}
f(p,x,t)&=1+\frac{\gam}{(z-\rho)c_{q,\gam}(\rho,x,t)}\psi_q(\rho,x,t)\Big( \psi_q(\rho,x,t)m_q(p,x,t)-\psi_q'(\rho,x,t)\Big)\\
&=1\mp\frac{\gam}{c_{q,\gam}(\rho,x,t)}\psi_q(\rho,x,t)^2\frac{1}{\I \sqrt{z}}+O(z^{-1}),\quad p=(z,\pm),
\end{align*}
where we have used that the Weyl--Titchmarsh $m$-function has the following asymptotic expansion for $p$ near $p_{\infty}$ (cf.~\cite[Lemma~6.1]{mikikits2009trace})
\begin{equation}
m_{q,\pm}(z,x,t)=\frac{\psi_{q,\pm}'(z,x,t)}{\psi_{q,\pm}(z,x,t)}=\pm\I \sqrt{z}+\frac{V_q(x,t)}{2\I \sqrt{z}}+O(z^{-1}), \quad p=(z,\pm).
\end{equation}
Thus comparing with \eqref{m2infp} proves the equation \eqref{equ:vminvqsol}.

To see uniqueness, let $\ti{m}_0(p)$ be a second solution which must be of the form
$\ti{m}_0(p) = \begin{pmatrix} \ti{f}(p^*) & \ti{f}(p) \end{pmatrix}$ by the symmetry condition.
Since the divisor $\dimu{x,t}$ is nonspecial, the Riemann--Roch theorem implies
$\ti{f}(p) = \alpha f(p) + \beta$ for some $\alpha,\beta\in \C$. But the pole condition implies
$\beta=0$ and the normalization condition implies $\alpha=1$.
\end{proof}

Since up to quasi-periodic factors $\psi_q(\rho,x,t)$ is a function of $x-v(\rho) t$, where
\be\label{equ:defvelosol}
v(\rho) = \frac{-12\re \int_{E_0}^{(\rho,+)}\! \omega_{p_{\infty},2}}{
\re \int_{E_0}^{(\rho,+)}\! \omega_{p_{\infty},0}},
\ee
we will call $v(\rho)$ the velocity of the corresponding soliton.

\section{The stationary phase points and the nonlinear dispersion relation}
\label{secSPP}

In this section we want to look at the relation between the energy $\lam$ of the underlying
Lax operator $H_q$ and the propagation speed at which the corresponding parts of KdV solutions travel, that is, the analog of the classical dispersion relation. If we set
\begin{equation} \label{def:v}
v(\lam) = \lim_{\eps\to 0} \frac{-12\, \re \big( \I \int_{E_0}^{(\lam+\I\eps,+)}\! \omega_{p_{\infty},2}\big)}{ \re \big( \I \int_{E_0}^{(\lam+\I\eps,+)}\! \omega_{p_{\infty},0}\big)}=
\lim_{\eps\to 0} \frac{-12\, \im \int_{E_0}^{(\lam+\I\eps,+)}\! \omega_{p_{\infty},2}}{ \im \int_{E_0}^{(\lam+\I\eps,+)}\! \omega_{p_{\infty},0}},
\end{equation}
our first aim is to show that the nonlinear dispersion relation is given by
\be
v(\lam)=\frac{x}{t}.
\ee
Recall that the Abelian differentials are given by \eqref{dsk0} and \eqref{dsk2}.

For $\rho\in\R\backslash\sig(H_q)$ the denominator is nonzero and the formula agrees with
the soliton velocity defined in \eqref{equ:defvelosol}. In particular, recalling the definition of our phase
$\phi$ from \eqref{defsp}, this implies
\be
v(\rho) = \frac{x}{t} \quad\Leftrightarrow\quad \re\,\phi(\rho,\frac{x}{t}) =0
\ee
in this case.
In particular, this definition reduces precisely to the definition of the velocity of a soliton
corresponding to the eigenvalue $\rho$ (cf. the discussion after Lemma~\ref{lem:singlesoliton}).

For $\lam\in\sig(H_q)$ both numerator and denominator vanish on $\sig(H_q)$ by \eqref{sgnR}. Hence by virtue of the de l'Hospital
rule we get
\be \label{equ:vinspectrum}
v(\lam) = -\frac{12\prod_{j=0}^g (\lam -\ti\lambda_j)}{\prod_{j=1}^g (\lam -\lambda_j)},
\ee
that is,
\be \label{equ:iffstatpoint}
v(\lam) = \frac{x}{t} \quad\Leftrightarrow\quad \phi'(\lam,\frac{x}{t}) =0.
\ee
In other words, $v(\lam)$ coincides with a stationary phase point in this case.

So let us discuss the stationary phase points, that is the solutions of $\phi'(\lam,\frac{x}{t}) =0$, next. The solutions
are given by the zeros of the polynomial
\be\label{stphpol}
12\prod_{j=0}^g (z -\tilde\lambda_j) + \frac{x}{t} \prod_{j=1}^g (z -\lambda_j) .
\ee
Since our Abelian differentials are all normalized to have vanishing $a_j$-periods, the numbers
$\lam_j$, $0\le j \le g$, are real and different with precisely one lying in each
spectral gap, say $\lam_j$ in the $j$'th gap.
Similarly, $\ti\lam_j$, $0\le j \le g$, are real and different and
$\ti\lam_j$, $1\le j \le g$, sits in the $j$'th gap. However $\ti\lam_0$ can be
anywhere (see \cite[Sect.~13.5]{tjac}).

The following lemma clarifies the dependence of the stationary phase points
on $x/t$.

\begin{lemma} \label{lem:statpoints}
Denote by $z_j(v)$, $0\le j \le g$, the stationary phase points, where $v=x/t$. Set $\lam_0=-\infty$ and $\lam_{g+1}= \infty$, then
\be
\lam_j < z_j(v) < \lam_{j+1}
\ee
and there is always at least one stationary phase point in each spectral gap.
Moreover, $z_j(v)$ is strictly decreasing with
\be
\lim_{v\to-\infty} z_j(v) = \lam_{j+1} \quad\text{and}\quad
\lim_{v\to\infty} z_j(v) = \lam_j.
\ee
\end{lemma}
\begin{proof}
Since the Abelian differential $\omega_{p_{\infty},2}+v \omega_{p_{\infty},0}$ has vanishing $a$ periods, the polynomial \eqref{stphpol}
must change sign in each gap except the lowest. Consequently there is at least one stationary phase
point in each gap except the lowest, and they are all different. Furthermore, by the implicit function theorem,
\[
z_j' = - \frac{q(z_j)}{\ti{q}'(z_j)+ v q'(z_j)} = -\frac{\prod_{k=1}^g (z_j -\lam_k)}{12\prod_{k=0,k\ne j}^g (z_j-z_k)},
\] 
where
\[
\ti{q}(z) = 12\prod_{k=0}^g (z -\tilde\lam_k), \quad
q(z)= \prod_{k=1}^g (z -\lam_k).
\]
Since the points $\lam_k$ are fixed points of this ordinary first order differential equation (note that the
denominator cannot vanish since the $z_j$'s are always different), the numbers
$z_j$ cannot cross these points. Combining the behavior as $v\to\pm\infty$ with the fact that
there must always be at least one of them in each gap, we conclude that $z_j$ must stay between
$\lam_j$ and $\lam_{j+1}$. This also shows $z_j' <0$ and thus $z_j(v)$ is strictly decreasing.
\end{proof}

In other words Lemma \ref{lem:statpoints} tells us the following: As $v =x/t$ runs from $-\infty$ to $\infty$ we start with $z_g(v)$ coming from $\infty$ towards $E_{2g}$, while the other stationary phase points $z_j$, $j=0,\dots ,g-1$, stay in their spectral gaps until $z_g(v)$ has passed $E_{2g}$ and therefore left the first spectral band $[E_{2g},\infty )$. After this has happened, the next stationary phase point $z_{g-1}(v)$ can leave its gap $(E_{2g-1},E_{2g})$ while $z_g(v)$ remains there, traverses the next spectral band $[E_{2g-2},E_{2g-1}]$ and so on. Finally $z_0(v)$ traverses the last spectral band $[E_0,E_1]$ and moves to $-\infty$.
So, depending on $x/t$ there is at most one single stationary phase point belonging to the union of the bands $\sigma(H_q)$ which is then 
the one solving \eqref{equ:iffstatpoint}.

\begin{lemma}
The function $v(\lam)$ defined in \eqref{def:v} is continuous and strictly decreasing.
Moreover, it is a bijection from $\R$ to $\R$.
\end{lemma}

\begin{proof}
That $v(\lam)$ defined in \eqref{def:v} is continuous is obvious except at the band edges $\lam=E_j$. However, in this case \eqref{def:v} becomes \eqref{equ:vinspectrum} by using the de l'Hospital rule. The function $v(\lam)$ defined in \eqref{equ:vinspectrum} is obviously continuous at the band edges $E_j$ since $\lam_j$ lies in the $j$'th gap and thus does not hit the band edges.

Furthermore, for large $\lam$ we have
\be
\lim_{\lam\to-\infty} \frac{v(\lam)}{-4\lam} =1, \quad
\lim_{\lam\to+\infty} \frac{v(\lam)}{-12\lam} =1,
\ee
which shows $\lim_{\lam\to\pm\infty} v(\lam) = \mp\infty$.

In the regions where there is one stationary phase point $z_j(v)\in \sigma(H_q)$ we know that $z_j(v)$
is the inverse of $v(\lam)$ and monotonicity follows from Lemma~\ref{lem:statpoints}. In the other regions we compute
\be
v'(z) = \frac{\prod_{j=0}^g (z -z_j(v(z)))}{-2\I \Rg{z} \int_{E_0}^{(z,+)}\!\! \omega_{p_{\infty},0}}.
\ee
Let the stationary phase points be ordered such that we have $z_j(v(z))=z$ for $z\in(E_{2j-1}-\eps,E_{2j-1}]$ and
$z_{j-1}(v(z))=z$ for $z\in[E_{2j},E_{2j}+\eps)$. Then we claim $z_j(v(z)) < z < z_{j-1}(v(z))$ for $z\in(E_{2j-1},E_{2j})$ (set $z_{-1}=\infty$).
In fact, the above differential equation implies that $v(z)$ can cross the curve $z_j(v(z))$ only from below and hence must stay
above this curve since it starts on this curve at $z=E_{2j-1}$. Similarly, it can cross the curve $z_{j-1}(v(z))$ only from below and would
hence remain above this curve afterwards. Thus this can only happen at $z=E_{2j}$.
\end{proof}

In summary, we can define a function $\zeta(x/t)$ via
\be\label{def:zeta}
v(\zeta)=\frac{x}{t}.
\ee
In particular, different solitons travel at different speeds
and don't collide with each other or the parts corresponding to
the continuous spectrum.

Moreover, there is some $\zeta_0$ for which $v(\zeta_0)=0$ and hence there can be {\em stationary solitons}
provided $\zeta_0\not\in\sig(H_q)$.

\begin{lemma}[Stationary solitons]
There exists a unique $\zeta_0$ such that $v(\zeta_0) = 0$. Moreover, if $\zeta_0 \in\sig(H_q)$ or $\ti{\lam}_0 \in\sig(H_q)$,
then $\zeta_0 = \ti{\lam}_0$. In particular, $\zeta_0\in\sig(H_q)$ if and only if $\ti{\lam}_0 \in\sig(H_q)$.
\end{lemma}

\begin{proof}
Existence and uniqueness of $\zeta_0$ follows since $v$ is a bijection.
It is left to show that $\zeta_0 = \ti{\lam}_0$ if $\zeta\in\sig(H_q)$ or $\ti{\lam}_0 \in\sig(H_q)$. Assume $\zeta_0\in \sig(H_q)$. Then using $v(\zeta_0)=0$ and \eqref{equ:vinspectrum} we get
\[
\prod_{j=0}^g(\zeta_0-\ti{\lam}_j)=(\zeta_0-\ti{\lam}_0)\prod_{j=1}^g(\zeta_0-\ti{\lam}_j)=0.
\]
Since $\ti{\lam}_j \in (E_{2j-1}, E_{2j})$, $j=1,\dots,g$ it follows $\zeta_0=\ti{\lam}_0$. Now suppose $\ti{\lam}_0\in \sig(H_q)$ and again use \eqref{equ:vinspectrum} to get
\[
v(\ti{\lam}_0)\prod_{j=1}^g(\ti{\lam}_0-\lam_j)=0.
\]
Since $\lam_j\in (E_{2j-1}, E_{2j})$, $j=1,\dots,g$ we obtain $v(\ti{\lam}_0)=0$ and thus $\zeta_0=\ti{\lam}_0$.
\end{proof}

In summary we conclude that depending on $v=\frac{x}{t}$ there can occur three cases:

\begin{enumerate} \label{casesregions}
\item
$\zeta(v) \in (E_{2j}, E_{2j+1})$ for some $j=0,\dots ,g$ (setting $E_{2g+1}=\infty$). In this case $\zeta(v)$ is a stationary phase point
and all other stationary phase points lie in open gaps.
\item
$\zeta(v) \in\R\setminus\sig(H_q)$ and all stationary phase points lie in open gaps.
\item
$\zeta(v) = E_j$ for some $j$ and all other stationary phase points lie in open gaps.
\end{enumerate}

These three cases define corresponding regions in the $(x,t)$-plane: the \emph{oscillatory region} (case (i)), the \emph{soliton region} (case (ii)), and the \emph{transitional region} (case (iii)).

\subsection*{Case (i): The oscillatory region}

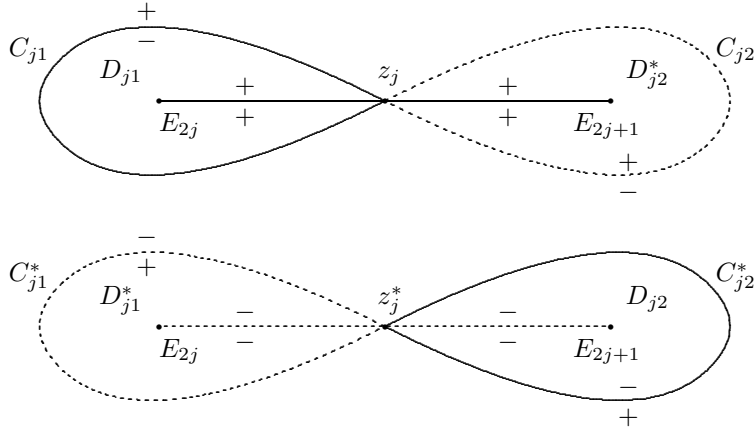
\begin{figure}
\begin{center}
\begin{picture}(10,6)

\put(1.2,4.8){$D_{j1}$}
\put(0,5.1){$C_{j1}$}
\put(8.2,4.8){$D_{j2}^*$}
\put(9.4,5.1){$C_{j2}$}
\put(5,4.5){\circle*{0.06}}
\put(4.9,4.8){$z_j$}
\put(2,4.5){\circle*{0.06}}
\put(2,4.1){$E_{2j}$}
\put(8,4.5){\circle*{0.06}}
\put(7.5,4.1){$E_{2j+1}$}

\put(3,4.6){$+$}
\put(3,4.2){$+$}
\put(6.5,4.6){$+$}
\put(6.5,4.2){$+$}

\put(1.7,5.6){$+$}
\put(1.7,5.2){$-$}
\put(8.1,3.6){$+$}
\put(8.1,3.2){$-$}

\put(2,4.5){\line(1,0){6}}
\put(5,4.5){\curve(0, 0, -2.645, -0.951, -4.28, -0.588, -4.28, 0.588, -2.645, 0.951, 0, 0)}
\curvedashes{0.05,0.05}
\put(5,4.5){\curve(0, 0, 2.645, -0.951, 4.28, -0.588, 4.28, 0.588, 2.645, 0.951, 0, 0)}
\curvedashes{}


\put(1.2,1.8){$D_{j1}^*$}
\put(0,2.1){$C_{j1}^*$}
\put(8.2,1.8){$D_{j2}$}
\put(9.4,2.1){$C_{j2}^*$}
\put(5,1.5){\circle*{0.06}}
\put(4.9,1.8){$z_j^*$}
\put(2,1.5){\circle*{0.06}}
\put(2,1.1){$E_{2j}$}
\put(8,1.5){\circle*{0.06}}
\put(7.5,1.1){$E_{2j+1}$}

\put(3,1.6){$-$}
\put(3,1.2){$-$}
\put(6.5,1.6){$-$}
\put(6.5,1.2){$-$}

\put(1.7,2.6){$-$}
\put(1.7,2.2){$+$}
\put(8.1,0.6){$-$}
\put(8.1,0.2){$+$}

\put(5,1.5){\curve(0, 0, 2.645, -0.951, 4.28, -0.588, 4.28, 0.588, 2.645, 0.951, 0, 0)}
\curvedashes{0.05,0.05}
\put(2,1.5){\curve(0,0,6,0)}
\put(5,1.5){\curve(0, 0, -2.645, -0.951, -4.28, -0.588, -4.28, 0.588, -2.645, 0.951, 0, 0)}

\end{picture}
\end{center}
\caption{The lens contour near a band containing a stationary phase point $z_j$ and its flipping image containing $z_j^*$.
Views from the top and bottom sheet. Dotted curves lie in the bottom sheet.\label{fig2}}
\end{figure}

Note that in this case we have
\be\label{phipp}
\phi''(z_j) / \I = -\frac{12 \prod_{k=0,k\ne j}^g (z_j -z_k)}{\I\Rg{z_j}} >0.
\ee
Suppose $\zeta(v)=z_j(v)$,  belongs to the interior of the band $[E_{2j}, E_{2j+1}]$ (with $E_{2g+1}=\infty$). Then we introduce the ``lens'' contour near that band as shown in Figure~\ref{fig2}.

The oriented paths $C_j = C_{j1} \cup C_{j2}$,  $C_j^* = C_{j1}^* \cup C_{j2}^*$ are meant to be close to the band $[E_{2j}, E_{2j+1}]$.

\begin{figure}
\begin{center}
\begin{picture}(10,6)

\put(3.4,4.8){$D_k$}
\put(0.6,5.15){$C_k$}
\put(2,4.5){\circle*{0.06}}
\put(2,4.1){$E_{2k}$}
\put(8,4.5){\circle*{0.06}}
\put(7.5,4.1){$E_{2k+1}$}

\put(4.8,4.6){$+$}
\put(4.8,4.2){$+$}

\put(5.5,5.6){$+$}
\put(5.5,5.2){$-$}

\put(2,4.5){\line(1,0){6}}
\put(5,4.5){\curve(0, 1, 2.645, 0.809, 4.28, 0.309, 4.28, -0.309, 2.645, -0.809, 0, -1, %
-2.645, -0.809, -4.28, -0.309, -4.28, 0.309, -2.645, 0.809, 0, 1)}


\put(3.4,1.8){$D_k^*$}
\put(0.6,2.2){$C_k^*$}
\put(2,1.5){\circle*{0.06}}
\put(2,1.1){$E_{2k}$}
\put(8,1.5){\circle*{0.06}}
\put(7.5,1.1){$E_{2k+1}$}

\put(4.8,1.6){$-$}
\put(4.8,1.2){$-$}

\put(5.5,2.6){$-$}
\put(5.5,2.2){$+$}

\curvedashes{0.05,0.05}
\put(2,1.5){\curve(0,0,6,0)}
\put(5,1.5){\curve(0, 1, 2.645, 0.809, 4.28, 0.309, 4.28, -0.309, 2.645, -0.809, 0, -1, %
-2.645, -0.809, -4.28, -0.309, -4.28, 0.309, -2.645, 0.809, 0, 1)}
\end{picture}
\end{center}
\caption{The lens contour near a band not including any stationary phase point. Views from the top and bottom sheet.\label{fig3}}
\end{figure}
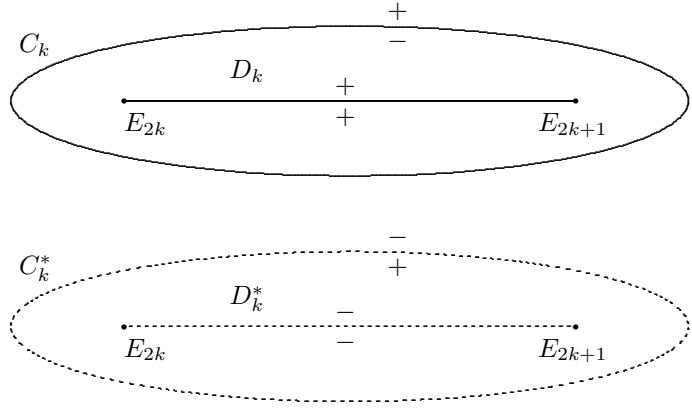

Concerning the  other bands $[E_{2k}, E_{2k+1}]$, $k\neq j$, $k=0,\dots,g$ (setting $E_{2g+1}=\infty$), one simply
constructs ``lens'' contours near each of the other bands $[E_{2k}, E_{2k+1}]$  and $[E_{2k}^*, E_{2k+1}^*]$  as shown in Figure~\ref{fig3}.

The oriented paths $C_k, C_k^*$ are meant to be  close to the band $[E_{2k}, E_{2k+1}]$. In particular, these loops must not
contain any of the eigenvalues $\rho_j$.

Then an investigation of the sign of $\re(\phi)$ shows the following:
\be
\begin{cases} \nn
\re(\phi(p) ) > 0 ,& p \in D_{j1}\cup D_k, \: \pi(p) < \zeta(x/t),\\
\re(\phi(p) ) < 0, & p \in D_{j2}\cup D_k, \: \pi(p) > \zeta(x/t),
\end{cases}
\ee
with $k=1,\dots,g$, $k\neq j$.

Observe that our original jump matrix \eqref{eq:jumpcond} has the following important factorization
\be \label{facb}
J(p)=b_-(p)^{-1} b_+(p),
\ee
where 
\be\label{defbt}
b_-(p)
=\begin{pmatrix}
1 &  R (p^*) \Theta(p^*)  \E^{-t\phi(p)} \\ 0 &1
\end{pmatrix}, \qquad
b_+(p)= \begin{pmatrix}
1 & 0 \\ R(p)  \Theta(p) \E^{t\phi(p)} & 1
\end{pmatrix},
\ee
which is the right factorization for $p\in \Sigma\backslash C(x/t)= \Sigma \cap \pi^{-1}((\zeta(x/t),\infty))$, i.e., $\pi(p)>\zeta(x/t)$. Similarly, we have
\be\label{facB}
J(p)= B_-(p)^{-1} \begin{pmatrix} 1-|R(p)|^2 & 0 \\ 0 & \frac{1}{1-|R(p)|^2}\end{pmatrix} B_+(p),
\ee
where 
\be\label{defBt}
B_-(p)
=\begin{pmatrix}
1 & 0 \\ - \frac{R(p)  \Theta(p)  \E^{t\phi(p)}}{1-|R(p)|^2} &1
\end{pmatrix}, \qquad
B_+(p)= \begin{pmatrix}
1 & -\frac{ R(p^*)  \Theta(p^*)  \E^{-t\phi(p)}}{1-|R(p)|^2} \\ 0 & 1
\end{pmatrix}.
\ee
This constitutes the right factorization for $p\in C(x/t)= \Sigma \cap \pi^{-1}((-\infty,\zeta(x/t))$, i.e., $\pi(p)<\zeta(x/t)$. Here we have used $\ol{R(p)}=R(p^*)$, for $p\in \Sigma$. To get rid of the diagonal part in the factorization corresponding to $\pi(p)<\zeta(x/t)$ and to conjugate the jumps near the eigenvalues we need to find the solution of the corresponding scalar Riemann--Hilbert problem, the so-called \emph{partial transmission coefficient}. Again we seek a meromorphic solution. This means that the poles of the scalar Riemann--Hilbert problem will be added to the resulting Riemann--Hilbert problem. On the other hand, a pole structure similar to the one of $m$ is crucial for uniqueness. We will address this problem by choosing the poles of the scalar problem in such a way that its zeros cancel the poles of $m$. The right choice will turn out to be $\dinuz$ (that is, the Dirichlet divisor corresponding to the limiting lattice defined in \eqref{limlatsol}).

Define a divisor $\dinu{x,t}$ of degree $g$ via
\be \label{defdinu}
\amap(\dinu{x,t}) = \amap(\dimu{x,t}) + \ul{\delta}(x/t),
\ee
where
\be \label{defdel}
\delta_\ell(x/t) = -2 \sum_{\rho_k<\zeta(x/t)} A_{E(\rho_k),\ell}(\rho_k) +
\frac{1}{2\pi\I} \int_{C(x/t)} \log(1-|R|^2) \zeta_\ell,
\ee
with $C(x/t) = \Sigma \cap \pi^{-1}((-\infty,\zeta(x/t))$ and $\zeta(x/t)$ as defined in \eqref{def:zeta}.

Then $\dinu{x,t}$ is nonspecial and $\pi(\hat{\nu}_j(x,t))=\nu_j(x,t)\in\R$ with precisely one
in each spectral gap (see \cite{kt2}).

We define the \emph{partial transmission coefficient} as
\be \label{defd}
\aligned
T(p,x,t) = & \frac{\theta \big( \ulz(p_{\infty},x,t)+\ul{\delta}(x/t)\big)}{\theta \big(\ulz(p_{\infty},x,t)\big)}
\frac{\theta \big(\ulz(p,x,t)\big)}{\theta \big(\ulz(p,x,t)+ \ul{\delta}(x/t)\big)} \cdot\\
& \cdot \bigg( \prod_{\rho_k<\zeta(x/t)} \!\!\! \exp\Big(-\int_{E_0}^p \om_{\rho_k\, \rho_k^*} \Big) \bigg)\exp\Big( \frac{1}{2\pi\I} \int_{C(x/t)} \log (1-|R|^2) \om_{p\,p_{\infty}}\Big),
\endaligned
\ee
where $\ul{\delta}(x,t)$ is defined in \eqref{defdel} and $\om_{p_1\,p_2}$
is the Abelian differential of the third kind with poles at $p_1$ and $p_2$.

The function $T(p,x,t)$ is meromorphic in $\mathcal{K}_g\setminus C(x/t)$ with first order poles at
$\rho_k<\zeta(x/t)$, $\hat{\nu}_j(x,t)$ and first order zeros at $\hat{\mu}_j(x,t)$.

\begin{lemma} \label{partialT}
$T(p,x,t)$ satisfies the following scalar meromorphic Riemann--Hilbert problem:
\be \label{rhpptc}
\aligned
&T_+(p,x,t) = T_-(p,x,t) (1-|R(p)|^2), \quad p \in C(x/t),\\
&(T(p,x,t))= \sum_{\rho_k<\zeta(x/t)} \di_{\rho_k^*} -\sum_{\rho_k<\zeta(x/t)} \di_{\rho_k} + \dimu{x,t}  -\dinu{x,t},\\
&T(p_{\infty},x,t) = 1.
\endaligned
\ee
Moreover,
\begin{enumerate}
\item
\[
T(p^*,x,t) T(p,x,t) = \prod_{j=1}^g \frac{z-\mu_j(x,t)}{z-\nu_j(x,t)}, \qquad z=\pi(p).
\]
\item
$\ol{T(p,x,t)}=T(\ol{p},x,t)$ and in particular $T(p,x,t)$ is real-valued for $\pi(p)\in\R\backslash\sig(H_q)$.
\end{enumerate}
\end{lemma}

\begin{proof}
The argument is similar to \cite[Thm.~4.3]{kt2}.
The solution of a Riemann--Hilbert problem on the Riemann sphere is given by the Plemelj-Sokhotsky formula. Since our problem is now set on the Riemann surface $\mathcal{K}_g$ the Cauchy kernel is given by the Abelian differential of the third kind $\omega_{p\,p_{\infty}}$ (cf.\ \cite{tag}). In particular,
$T(p,x,t)$ satisfies the jump condition from \eqref{rhpptc} along $C(x/t)$. Next we have to check that the function $T(p,x,t)$ extends to a single-valued function on $\mathcal{K}_g$. For that purpose note that the only possible contribution which causes multi-valuedness may come from the $b$-cycles since all Abelian differentials are normalized to have vanishing $a$-periods. So for the $b_{\ell}$-periods $\ell=1,\dots,g$ we compute for $p\in C(x/t)$
\[
\lim_{\eps\downarrow 0} \frac{T(p + i \eps,x,t)}{T(p - i \eps,x,t)} = \exp\Big( 2\pi\I \delta_{\ell}-\int_{C(x/t)}\log (1-|R|^2)\zeta_{\ell}+
\sum_{\rho_k<\zeta(x/t)}4\pi\I A_{E(\rho_k),\ell}(\rho_k))\Big),
\]
which is indeed equals $1$ by the choice of $\delta_{\ell}$ in \eqref{defdel}. 

Concerning the poles and zeros of the function $T(p,x,t)$ we see that by Riemann's vanishing theorem (cf.~\cite[Theorem~A.13]{tjac}) and the choice of the divisor $\dinu{x,t}$ defined by \eqref{defdinu} that the ratio of theta functions is meromorphic with simple zeros at $\hat{\mu}_j$ and simple poles at $\hat{\nu}_j$. Moreover, from the product of the Blaschke factors we get that $T$ has simple poles at $\rho_k$ and simple zeros at $\rho_k^*$ for which $\rho_k<\zeta(x/t)$ is valid.

To prove uniqueness let $\ti{T}$ be a second solution and consider $\ti{T}/T$. Then
$\ti{T}/T$ has no jump and the Schwarz reflection principle implies that it extends
to a meromorphic function on $\mathcal{K}_g$. Since the poles of $T$ cancel the poles of $\ti{T}$,
its divisor satisfies $(\ti{T}/T) \ge -\dimu{x,t}$. Since $\dimu{x,t}$ is nonspecial, $\ti{T}/T$ has to be a constant by the Riemann--Roch theorem (cf.~\cite[Theorem~A.2]{tjac}). Setting $p=p_{\infty}$,
we see that this constant is one, that is, $\ti{T}=T$ as claimed.

Finally, $\ol{T(p,x,t)}=T(\ol{p},x,t)$ follows from uniqueness since both functions solve \eqref{rhpptc}.
\end{proof}

We will also need the expansion around $p_{\infty}$ given by
\begin{lemma}
\label{thm:Tinfp}
The asymptotic expansion of the partial transmission coefficient for $p$ near $p_{\infty}$ is given by
\be\label{Tinfp}
T(p,x,t) = 1 \pm \frac{T_1(x,t)}{\sqrt{z}} + O(\frac{1}{z}), \qquad p=(z,\pm),
\ee
where
\be \label{eq:T1exp}
\aligned
T_1(x,t) =& -\sum_{\rho_k<\zeta(x/t)} 2\int_{E(\rho_k)}^{\rho_k} \omega_{p_{\infty},0}
+ \frac{1}{2\pi\I} \int_{C(x/t)} \log(1-|R|^2) \omega_{p_{\infty},0} \\
& {} - \I \partial_y \ln \left( \frac{\theta \big( \ulz(p_{\infty},y,t) + \ul{\delta}(x/t)\big)}{\theta\big( \ulz(p_{\infty},y,t)\big)} \right) \Bigg|_{y=x},
\endaligned
\ee
and $\omega_{p_{\infty},0}$ is the Abelian differential of the second kind defined in \eqref{dsk0}.
\end{lemma}

\begin{proof}
This can be verified similarly as in the case of the full transmission coefficient (cf.~\cite[Theorems~6.2 and 6.3]{mikikits2009trace}
and by expanding the ratio of theta functions near $p_{\infty}$. 
\end{proof}

Now that we have solved the scalar Riemann--Hilbert problem for $T(p,x,t)$ we can conjugate our original Riemann--Hilbert problem.

Since to each discrete eigenvalue there corresponds a soliton, it follows that solitons are represented in our Riemann--Hilbert problem
by the pole conditions \eqref{eq:jumpcond2}. For this reason we will study how poles can be dealt with in this section.
We will follow closely the presentation of \cite[Section~4]{krt}.

In order to remove the poles there are two cases to distinguish. If $\rho_j > \zeta(x/t)$, the jump
at $\rho_j$ is exponentially close to the identity and there is nothing to do. 

Otherwise, if $\rho_j < \zeta(x/t)$, we need to use conjugation to turn the jumps at these poles into exponentially decaying ones, following \cite{dkkz}. It turns out that we will have to handle the poles at $\rho_j$ and $\rho_j^*$ in one step in order to preserve symmetry and in order to not add additional poles elsewhere.

Moreover, the conjugation of the Riemann--Hilbert problem also serves another purpose, namely that the jump matrix can be separated into two matrices, one containing an off-diagonal term with $\exp(-t\phi)$ and the other with $\exp(t\phi)$. Without conjugation this is not possible for the jump on $C(x/t)=\Sigma\cap \pi^{-1}\big(  (-\infty,\zeta(x/t))\big)$, since in this case there also appears a diagonal matrix if one wants to separate the jump matrix.

For easy reference we note the following result, which can be verified by a straightforward calculation.
\begin{lemma}[Conjugation]\label{lem:conjug}
Assume that $\widetilde{\Sigma}\subseteq\Sigma$. Let $D$ be a matrix of the form
\be
D(p) = \begin{pmatrix} d(p^*) & 0 \\ 0 & d(p) \end{pmatrix},
\ee
where $d: \mathcal{K}_g\backslash\widetilde{\Sigma}\to\C$ is a sectionally analytic function. Set
\be
\ti{m}(p) = m(p) D(p),
\ee
then the jump matrix transforms according to
\be
\ti{J}(p) = D_-(p)^{-1} J(p) D_+(p).
\ee
$\ti{m}(p)$ satisfies the symmetry condition \eqref{eq:symcond} if and only if $m(p)$ does.
Furthermore, $\ti{m}(p)$ satisfies the normalization condition \eqref{eq:normcond} if $m(p)$
satisfies \eqref{eq:normcond} and $d(p_{\infty})= 1$.
\end{lemma}

\begin{lemma}[\cite{krt3}, Lem.~7.2]\label{lem:intib}
Introduce
\be
\ti{B}(p,\rho) =
C_\rho(x,t)
\frac{\theta(\ul{z}(p,x,t))}{\theta(\ul{z}(p,x,t) + 2 \Amap(\rho))} B(p,\rho).
\ee
Then $\ti{B}(.,\rho)$ is a well defined meromorphic function, with divisor
\be
(\ti{B}(.,\rho)) = -\mathcal{D}_{\hat{\ul{\nu}}} + \mathcal{D}_{\hat{\ul{\mu}}}
- \mathcal{D}_{\rho^*} + \mathcal{D}_{\rho},
\ee
where $\nu$ is defined via
\be
\ul{\alpha}_{E_0}(\mathcal{D}_{\hat{\ul{\nu}}}) = 
\ul{\alpha}_{E_0}(\mathcal{D}_{\hat{\ul{\mu}}}) + 2 \Amap(\rho).
\ee
Furthermore,
\be\label{eq:normtib}
\ti{B}(p_{\infty},\rho)= 1,
\ee
if
\be
C_\rho(x,t) = \frac{\theta(\ul{z}(p_{\infty},x,t) + 2 \Amap(\rho))}{\theta(\ul{z}(p_{\infty},x,t))}.
\ee
\end{lemma}

Now, we can show how to conjugate the jump corresponding to one eigenvalue following \cite{krt3}.

\begin{lemma}\label{lem:twopolesinc}
Assume that the Riemann--Hilbert problem for $m$ has jump conditions near $\rho$ and
$\rho^*$ given by
\be
\aligned
m_+(p)&=m_-(p)\begin{pmatrix}1& 0 \\ \frac{\gam(p)}{\pi(p) - \rho} &1\end{pmatrix}, \qquad p\in\Sigma_\eps(\rho),\\
m_+(p)&=m_-(p)\begin{pmatrix}1& -\frac{\gam(p^*)}{\pi(p) - \rho} \\ 0 &1\end{pmatrix}, \qquad p\in\Sigma_\eps(\rho^*),
\endaligned
\ee
and satisfies a divisor condition 
\be
(m_1) \ge -\dimuzs, \qquad (m_2) \ge -\dimuz.
\ee
Then this Riemann--Hilbert problem is equivalent to a Riemann--Hilbert problem for $\ti{m}$
which has jump conditions near $\rho$ and $\rho^*$ given by
\be
\aligned
\ti{m}_+(p)&= \ti{m}_-(p)\begin{pmatrix}1& \frac{\ti{B}(p,\rho^*) (\pi(p) - \rho)}{\gam(p) \ti{B}(p^*,\rho^*)} \\ 0 &1\end{pmatrix},
\qquad p\in\Sigma_\eps(\rho),\\
\ti{m}_+(p)&= \ti{m}_-(p)\begin{pmatrix}1& 0 \\ -\frac{\ti{B}(p^*,\rho^*) (\pi(p) - \rho)}{\gam(p^*) \ti{B}(p,\rho^*)} &1\end{pmatrix},
\qquad p\in\Sigma_\eps(\rho^*),
\endaligned
\ee
divisor condition
\be
(\ti{m}_1) \ge -\dinuzs, \qquad (\ti{m}_2) \ge -\dinuz,
\ee
where $\dinuz$ is defined via
\be
\ul{\alpha}_{E_0}(\mathcal{D}_{\hat{\ul{\nu}}}) = 
\ul{\alpha}_{E_0}(\mathcal{D}_{\hat{\ul{\mu}}}) + 2 \Amap(\rho),
\ee
and all remaining data conjugated (as in Lemma~\ref{lem:conjug}) by
\be
D(p) = \begin{pmatrix} \ti{B}(p^*,\rho^*) & 0 \\ 0 & \ti{B}(p,\rho^*) \end{pmatrix}.
\ee
\end{lemma}

\begin{proof}
Denote by $U$ the interior of $\Sigma_\eps(\rho)$.
To turn $\gam$ into $\gam^{-1}$, introduce $D$ by
\[
D(p) = \begin{cases}
\begin{pmatrix} 1 & \frac{\pi(p) - \rho}{\gam(p)} \\
-\frac{\gam(p)}{\pi(p) - \rho} & 0 \end{pmatrix}
\begin{pmatrix} \ti{B}(p^*,\rho^*) & 0 \\ 0 & \ti{B}(p,\rho^*) \end{pmatrix}, &  p \in U, \\
\begin{pmatrix} 0 & -\frac{\gam(p^*)}{\pi(p) - \rho} \\ \frac{\pi(p) - \rho}{\gam(p^*)} & 1 \end{pmatrix}
\begin{pmatrix} \ti{B}(p^*,\rho^*) & 0 \\ 0 & \ti{B}(p,\rho^*) \end{pmatrix}, & p^* \in U, \\ 
\begin{pmatrix} \ti{B}(p^*,\rho^*) & 0 \\ 0 & \ti{B}(p,\rho^*) \end{pmatrix}, & \text{else},
\end{cases}
\]
and note that $D(p)$ is meromorphic away from the two circles. Now set $\ti{m}(p) = m(p) D(p)$.
The claim about the divisors follows from noting, where the poles of $\ti{B}(p,\rho)$ are.
\end{proof}

Note that Lemma~\ref{lem:twopolesinc} can be applied iteratively to conjugate the eigenvalues $\rho_j<\zeta(x/t)$: start with the poles $\mu=\mu^0$ and apply the lemma setting $\rho=\rho_1$. This results in new poles $\mu^1=\nu$. Then repeat this with $\mu=\mu^1$, $\rho=\rho_2$, and so on.

All in all we will now make the following conjugation step: abbreviate
\[
\gam_k(p,x,t)= 
\frac{-2 \Rg{\rho_k}}{\prod_{l=1}^g (\rho_k-\mu_l)} \frac{\psi_q(p,x,t)}{\psi_q(p^*,x,t)} \gam_k
\]
and introduce
\be \label{def:D}
D(p) = \begin{cases}
\begin{pmatrix} 1 & \frac{\pi(p) - \rho_k}{\gam_k(p,x,t)} \\ -\frac{\gam_k(p,x,t)}{\pi(p) - \rho_k} & 0 \end{pmatrix}
D_0(p), & \begin{smallmatrix}|\pi(p)-\rho_k|<\eps\\ p\in\Pi_+\end{smallmatrix}, \: \rho_k < \zeta(x/t), \\
\begin{pmatrix} 0 & -\frac{\gam_k(p^*,x,t)}{\pi(p) - \rho_k} \\ \frac{\pi(p) - \rho_k}{\gam_k(p^*,x,t)} & 1 \end{pmatrix}
D_0(p), & \begin{smallmatrix}|\pi(p)-\rho_k|<\eps\\ p\in\Pi_-\end{smallmatrix}, \: \rho_k < \zeta(x/t),\\ 
D_0(p), & \text{else},
\end{cases}
\ee
where 
\[
D_0(p) = \begin{pmatrix} T(p^*,x,t) & 0 \\ 0 & T(p,x,t) \end{pmatrix}.
\]
Note that $D(p)$ is meromorphic in $\mathcal{K}_g\backslash C(x/t)$ and that we have
\be \label{eq:symforD}
D(p^*)= \sigI D(p) \sigI.
\ee
Now we conjugate our problem using $D(p)$:

\begin{theorem}[Conjugation] \label{thm:conjugation}
The function $m^2(p)=m(p)D(p)$, where $D(p)$ is defined in \eqref{def:D}, is meromorphic away from $C(x/t)$ and satisfies:
\begin{enumerate}
\item The jump condition
\be
m^2_+(p) = m^2_-(p) J^2(p), \quad p\in\Sigma, \label{rhpm4}
\ee
where the jump matrix is given by
\be \label{def:j2}
J^2(p) = D_{0-}(p)^{-1} J(p) D_{0+}(p),
\ee
\item the divisor conditions
\be \label{divm2}
(m^2_1) \ge -\dinus{x,t}, \quad (m^2_2) \ge -\dinu{x,t},
\ee
All jumps corresponding to poles, except for possibly one if
$\rho_k=\zeta(x/t)$, are exponentially decreasing. In that case we will keep the pole condition which is now of the form:
\be
\aligned \label{eq:specpolecond}
& \Big( m^2_1(p) + \frac{\gam_k(p,x,t)}{\pi(p) - \rho_k}\frac{T(p^*,x,t)}{T(p,x,t)} m^2_2(p) \Big) \ge - \dinus{x,t},
\mbox{ near $\rho_k$},\\
& \Big( \frac{\gam_k(p^*,x,t)}{\pi(p) - \rho_k} \frac{T(p,x,t)}{T(p^*,x,t)}  m^2_1(p) + m^2_2(p) \Big) \ge - \dinu{x,t},
\mbox{ near $\rho_k^*$}.
\endaligned
\ee 
\item the symmetry condition
\[
m^2(p^*) = m^2(p) \sigI,
\]
\item and the normalization
\[
m^2(p_{\infty }) = \begin{pmatrix}  1 & 1 \end{pmatrix}.
\]
\end{enumerate}
\end{theorem}
\begin{proof}
Invoking Lemma~\ref{lem:conjug} and \eqref{eq:jumpcond} we see that the jump matrix $J^2(p)$ is indeed given by \eqref{def:j2}. The divisor conditions follow from the one for $T(p,x,t)$ and $m(p)$.
Moreover, using Lemma~\ref{lem:twopolesinc} one easily sees that the jump corresponding to $\rho_k < \zeta(x/t)$ (if any) is given by
\be
\aligned
J^2(p) &= \begin{pmatrix}1& \frac{T(p,x,t) (\pi(p) - \rho_k)}{\gam_k(p,x,t) T(p^*,x,t)} \\ 0 &1\end{pmatrix},
\qquad p\in\Sigma_\eps(\rho_k),\\
J^2(p) &= \begin{pmatrix}1& 0 \\ -\frac{T(p^*,x,t) (\pi(p) - \rho_k)}{\gam_k(p^*,x,t) T(p,x,t)} &1\end{pmatrix},
\qquad p\in\Sigma_\eps(\rho_k^*),
\endaligned
\ee
and by Lemma~\ref{lem:conjug} the jump corresponding to $\rho_k > \zeta(x/t)$ (if any) reads
\be
\aligned
J^2(p) &= 
\begin{pmatrix} 1 & 0 \\
\frac{\gam_k(p,x,t) T(p^*,x,t)}{T(p,x,t) (\pi(p)-\rho_k)} & 1 \end{pmatrix},
\qquad p\in\Sigma_\eps(\rho_k),\\
J^2(p) &= 
\begin{pmatrix} 1 & -\frac{\gam_k(p^*,x,t) T(p,x,t)}{T(p^*,x,t) (\pi(p)-\rho_k)} \\ 
0 & 1 \end{pmatrix},
\qquad p\in\Sigma_\eps(\rho_k^*).
\endaligned
\ee
That is, all jumps corresponding to the poles $\rho_k\neq \zeta(x/t)$ are exponentially decreasing.
That the pole conditions are of the form \eqref{eq:specpolecond} in the case $\rho_k=\zeta(x/t)$ can be checked directly: just use the pole conditions of the original Riemann--Hilbert problem \eqref{eq:polecond} and the divisor condition \eqref{rhpptc} for $T(p,x,t)$.
Furthermore, by \eqref{eq:symcond} and \eqref{eq:symforD} one checks that the symmetry condition for $m^2$ is fulfilled.
From $T(p_{\infty},x,t)=1$ we finally deduce
\be
m^2(p_{\infty })=m(p_{\infty })=\begin{pmatrix}  1 & 1 \end{pmatrix},
\ee
which finishes the proof.
\end{proof}

For $p\in\Sigma\setminus C(x/t)=\Sigma\cap \pi^{-1}\big( (\zeta(x/t),\infty)\big)$ the jump matrix $J^2$ can be factorized as
\be \nn
J^2 = (\ti{b}_-)^{-1} \ti{b}_+,
\ee
where $\ti{b}_\pm = D_0^{-1} b_\pm D_0$, that is,
\be \label{equ:bpmtidef}
\ti{b}_- = 
\begin{pmatrix}
1 & \frac{T(p,x,t)}{T(p^*,x,t)}R(p^*) \Theta(p^*) \E^{-t\phi(p)} \\
0 & 1 \end{pmatrix},\quad
\ti{b}_+ =
\begin{pmatrix}
1 & 0 \\
\frac{T(p^*,x,t)}{T(p,x,t)} R(p) \Theta(p) \E^{t\phi(p)} & 1 \end{pmatrix}.
\ee
For $p\in C(x/t)=\Sigma\cap \pi^{-1}\big( (-\infty,\zeta(x/t))\big)$ we can factorize $J^2$ in the following way
\be \nn
J^2 = (\ti{B}_-)^{-1} \ti{B}_+,
\ee
where $\ti{B}_\pm = D_{\pm}^{-1} B_\pm D_{\pm}$, that is,
\be \label{equ:Bpmtidef}
\ti{B}_- =\begin{pmatrix} 1 & 0 \\-\frac{T_-(p^*,x,t)}{T_-(p,x,t)}  \frac{R(p)  \Theta(p)}{1-|R(p)|^2} \E^{t\, \phi(p)} & 1\end{pmatrix},\quad
\ti{B}_+ =\begin{pmatrix} 1 & - \frac{T_+(p,x,t)}{T_+(p^*,x,t)} \frac{R(p^*) \Theta(p^*)}{1-|R(p)|^2} \E^{-t\, \phi(p)} \\0 & 1 \end{pmatrix}.
\ee

Note that by $\ol{T(p,x,t)}=T(\ol{p},x,t)$ we have
\be
\frac{T_-(p^*,x,t)}{T_+(p,x,t)} = \frac{T_-(p^*,x,t)}{T_-(p,x,t)} \frac{1}{1-|R(p)|^2} =
\frac{\ol{T_+(p,x,t)}}{T_+(p,x,t)}, \quad p\in C(x/t),
\ee
respectively
\be
\frac{T_+(p,x,t)}{T_-(p^*,x,t)} = \frac{T_+(p,x,t)}{T_+(p^*,x,t)} \frac{1}{1-|R(p)|^2} =
\frac{\ol{T_-(p^*,x,t)}}{T_-(p^*,x,t)} , \quad p\in C(x/t).
\ee

We are now able to redefine the Riemann--Hilbert problem for $m^2(p)$ in such a way that the jumps of the new Riemann--Hilbert problem will lie in the regions where they are exponentially close to the identity for large times.
The following theorem can be checked by straightforward calculations:
\begin{theorem}[Deformation]
Define $m^3(p)$ by
\be \label{defm5}
m^3(p) = \begin{cases}
m^2(p) \ti{B}_+(p)^{-1}, & p \in D_k\cup D_{j1}, \: k < j,\\
m^2(p) \ti{B}_-(p)^{-1}, & p \in D_k^*\cup D_{j1}^*, \: k < j,\\
m^2(p) \ti{b}_+(p)^{-1}, & p \in D_k\cup D_{j2}, \: k > j,\\
m^2(p) \ti{b}_-(p)^{-1}, & p \in D_k^*\cup D_{j2}^*, \: k > j,\\
m^2(p), & \text{else},
\end{cases}
\ee
where the matrices $\ti{b}_\pm$ and $\ti{B}_\pm$ are defined in \eqref{equ:bpmtidef} and \eqref{equ:Bpmtidef}, respectively. Here we assume that the deformed contour is sufficiently close to
the original one. Then the function $m^3(p)$ satisfies:
\begin{enumerate}
\item The jump condition
\be
m^3_+(p) = m^3_-(p) J^3(p), \mbox{ for $p\in\Sigma$},
\ee
where the jump matrix $J^3$ is given by
\be \label{rhpm5jump}
J^3(p) = \begin{cases}
\ti{B}_+(p), & p \in C_k\cup C_{j1}, \: k < j,\\
\ti{B}_-(p)^{-1}, & p \in C_k^*\cup C_{j1}^*, \: k < j,\\
\ti{b}_+(p), & p \in C_k\cup C_{j2}, \: k > j,\\
\ti{b}_-(p)^{-1}, & p \in C_k^*\cup C_{j2}^*, \: k > j,\\
J^2(p), & \text{else},
\end{cases}
\ee
\item the divisor conditions
\be
(m^3_1) \ge -\dinus{x,t}, \quad (m^3_2) \ge -\dinu{x,t},
\ee
The jumps on the small circles around the eigenvalues remain unchanged.
\item The symmetry condition
\be
m^3(p^*) = m^3(p) \sigI,
\ee
\item
and the normalization
\be
m^3(p_{\infty}) =\begin{pmatrix} 1 & 1\end{pmatrix}.
\ee
\end{enumerate}
\end{theorem}

Here we have assumed that the reflection coefficient $R(p)$ appearing in the jump matrices admits an analytic extension to the corresponding regions. Of course this is not true in general, but we can always evade this obstacle by approximating $R(p)$ by analytic functions. We relegate the details to Section~\ref{secanap}.

The crucial observation now is that the jumps  $J^3$ on the oriented paths $C_k$, $C_k^*$ are of the form $\id+ exponentially~small$ asymptotically as $t \to \infty$, at least away from the stationary phase points $z_j$, $z^*_j$. We thus hope we can simply replace these jumps by the identity matrix (asymptotically as $t \to \infty$) implying that the solution should asymptotically be given by the constant vector $\rI$. That this can in fact be done will be shown in the next section by explicitly computing the contribution of the stationary phase points thereby showing that they are of the order $O(t^{-1/2})$, that is,
\[
m^3(p) = \rI + O(t^{-1/2})
\]
uniformly for $p$ away from the jump contour. Hence all which remains to obtain the leading term $V_l$ in Theorem~\ref{thmMain3}
is to trace back the definitions
of  $m^3$ and $m^2$ and comparing with \eqref{m2infp}. First of all, since $m^3$ and $m^2$ coincide near $p_{\infty}$ we have
\[
m^2(p) = \rI + O(t^{-1/2})
\]
uniformly for $p$ in a neighborhood of $p_{\infty}$. Consequently, by the definition of $m^2$ (see Theorem~\ref{thm:conjugation}), we have
\[
m(p) = \begin{pmatrix} T(p^*,x,t)^{-1} & T(p,x,t)^{-1} \end{pmatrix} + O(t^{-1/2})
\]
again uniformly for $p$ in a neighborhood of $p_{\infty}$. Finally, using the expansion of $T(p,x,t)$ near $p_{\infty}$ (see Lemma~\ref{thm:Tinfp}) and then comparing the last identity with \eqref{m2infp} shows
\be
\int_x^{\infty}(V-V_q)(y,t)dy=2\I T_1(x,t)+O(t^{-1/2}),
\ee
where $T_1$ is defined via \eqref{eq:T1exp}, that is,
\begin{align*}
T_1(x,t) =& -\sum_{\rho_k<\zeta(x/t)} 2\int_{E(\rho_k)}^{\rho_k} \omega_{p_{\infty},0}
+ \frac{1}{2\pi\I} \int_{C(x/t)} \log(1-|R|^2) \omega_{p_{\infty},0} \\
& {} - \I \partial_x \ln \left( \frac{\theta \big( \ulz(p_{\infty},x,t) + \ul{\delta}(x/t)\big)}{\theta\big( \ulz(p_{\infty},x,t)\big)} \right) .
\end{align*}
Similarly one obtains
\be
(V-V_q)(x,t)= O(t^{-1/2}),
\ee
by using \eqref{equ:expm1m2} instead of \eqref{m2infp}.

Hence we have proven the leading term in Theorem~\ref{thmMain3}, the next term will be computed in Section~\ref{secCROSS}.

\subsection*{Case (ii): The soliton region}

In the case where no stationary phase points lie in the spectrum the situation is similar to the case~(i). In fact, it is much simpler since there is no contribution from the stationary phase points: There is a gap (the $j$-th gap, say) in which two stationary phase points exist. Similarly as in case~(i) an investigation of the sign of $\re(\phi)$ shows the following:
\be
\begin{cases} \nn
\re(\phi(p) ) > 0 ,& p \in D_k, \: k < j,\\
\re(\phi(p) ) < 0, & p \in D_k, \: k > j.
\end{cases}
\ee
Now we construct ``lens-type'' contours $C_k$ (as shown in Figure~\ref{fig3}) around every single band lying to the left of the $j$-th gap and make use of the factorization $J^2=(\ti{b}_-)^{-1} \ti{b}_+$, where the matrices $\ti{b}_-$ and $\ti{b}_+$ are defined in \eqref{equ:bpmtidef}. We also construct such ``lens-type'' contours $C_k$ around every single band lying to the right of the $j$-th gap and make use of the factorization $J^2=(\ti{B}_-)^{-1} \ti{B}_+$ with the matrices $\ti{B}_-$ and $\ti{B}_+$ given by \eqref{equ:Bpmtidef}. Indeed, in place of \eqref{defm5}
we set
\be
m^3(p) = \begin{cases} m^2(p)  \ti{B}_+^{-1}(p), & p \in D_k, \: k<j,\\
m^2(p)  \ti{B}_-^{-1}(p), & p \in D_k^*, \: k<j,\\
m^2(p)  \ti{b}_+^{-1}(p), & p \in D_k, \: k>j,\\
m^2(p)  \ti{b}_-^{-1}(p), & p \in D_k^*, \: k>j,\\
m^2(p), & \text{else}.
\end{cases}
\ee

Now we are ready to prove Theorem~\ref{thmMain2} by applying Theorem~\ref{thm:remcontour} in the following way:

If $|\zeta(x/t) - \rho_k|>\eps$ for all $k$ we can choose $\gam_0^t=0$ and $w_0^t$ by removing all jumps
corresponding to poles from $w^t$. The error between the solutions of $w^t$ and $w_0^t$
is exponentially small in the sense of Theorem~\ref{thm:remcontour}, that is, $\| w^t -w_0^t\|_{\infty}\leq O(t^{-l})$ for any $l\geq 1$. We have the one soliton solution (cf.\ Lemma~\ref{lem:singlesoliton}) $\hat{m}_0(p) = \begin{pmatrix} \hat{f}(p^*,x,t) & \hat{f}(p,x,t) \end{pmatrix}$, where $\hat{f}(p)=1$ for $p$ large enough. Using Lemma~\ref{thm:Tinfp} we compute
\begin{align*}
m(p) =& \hat{m}_0(p) \begin{pmatrix} T(p^*,x,t)^{-1} & 0\\ 0 & T(p,x,t)^{-1} \end{pmatrix}\\
=&\begin{pmatrix} 1+\frac{T_1(x,t)}{\sqrt{z}}+O(z^{-1}) & 1-\frac{T_1(x,t)}{\sqrt{z}}+O(z^{-1})\end{pmatrix}.
\end{align*}
Comparing this expression with \eqref{m2infp} yields
\[
\int_x^{\infty}(V-V_q)(y,t)dy=2\I T_1(x,t)+O(t^{-n}),
\]
and thus by our definition of the limiting solution we finally have
\[
\int_x^{\infty}(V-V_l)(y,t)dy=\int_x^{\infty}\Big( (V-V_q)(y,t)-(V_l-V_q)(y,t)\Big)dy=O(t^{-n}),
\]
for any $n\geq1$ if $R(p)$ has an analytic extension. This proves the second part of the theorem.

If $|\zeta(x/t) - \rho_k|<\eps$ for some $k$, we choose $\gam_0^t=\tilde{\gam }_k$ and $w_0^t \equiv 0$.
Again we conclude that the error between the solutions of $w^t$ and $w_0^t$ is exponentially small, that is, $\| w^t -w_0^t\|_{\infty}\leq O(t^{-l})$, for any $l \geq 1$. By Lemma~\ref{lem:singlesoliton} we have the one soliton solution $\hat{m}_0(p) = \begin{pmatrix} \hat{f}(p^*,x,t) & \hat{f}(p,x,t) \end{pmatrix}$, with 
\[
\hat{f}(p,x,t)=1+\frac{\tilde{\gam}_k}{z-\rho_k} \frac{\psi_{l,c_k}(\rho_k,x,t)
W_{(x,t)}(\psi_{l,c_k}(\rho_k,x,t),\psi_{l,c_k}(p,x,t))}{\psi_{l,c_k}(p,x,t)c_{l,k}(x,t)},
\]
for $p$ large enough, where $\tilde{\gam}_k$ is defined as in \eqref{eq:gamshift}. We will again use
\[
m(p) = \hat{m}_0(p) \begin{pmatrix} T(p^*,x,t)^{-1} & 0\\ 0 & T(p,x,t)^{-1} \end{pmatrix}=
\begin{pmatrix} \frac{\hat{f}(p^*,x,t)}{T(p^*,x,t)} & \frac{\hat{f}(p,x,t)}{T(p,x,t)}\end{pmatrix},
\]
and now expand $\hat{f}(p)$ as in the proof of Lemma~\ref{lem:singlesoliton}. Finally a comparison with \eqref{m2infp} yields
\[
\int_x^{\infty}(V-V_q)(y,t)dy=2\I T_1(x,t) +2\frac{\tilde{\gam}_k\psi_{l,c_k}(\rho_k,x,t)^2}{c_{l,k}(x,t)}+O(t^{-n}),
\]
and hence by our definition of the limiting solution \eqref{limlatsol} we obtain \eqref{eq:sol1} for any $n\geq1$ if $R(p)$ has an analytic extension.
Similarly we obtain \eqref{eq:sol2} by using \eqref{equ:expm1m2} instead of \eqref{m2infp}.

\subsection*{Case (iii): The transitional region}

In the  special case where the two stationary phase points coincide
(so $z_j = z_j^*= E_k$ for some $k$) the  Riemann--Hilbert problem arising above is of a different nature.
We expect a similar behavior as in the constant background case \cite{dvz} (cf.\ also the discussion in \cite{kt2}).
However, we will not treat this case here.

\section{The ``local'' Riemann--Hilbert problems on the small crosses} \label{secCROSS}

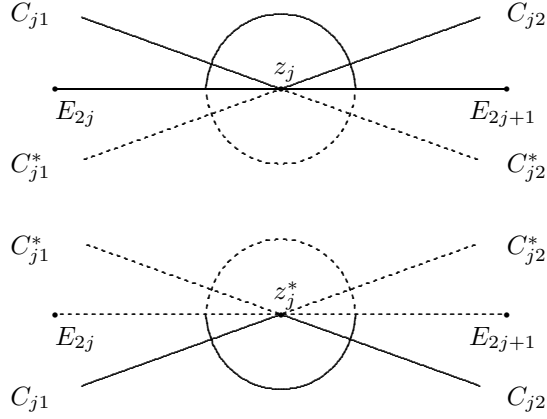
\begin{figure}
\begin{center}
\begin{picture}(10,6)

\put(1.4,5.4){$C_{j1}$}
\put(8.0,5.4){$C_{j2}$} 
\put(1.4,3.4){$C_{j1}^*$} 
\put(8.0,3.4){$C_{j2}^*$}  
\put(5,4.5){\circle*{0.06}}
\put(4.9,4.7){$z_j$}
\put(2,4.5){\circle*{0.06}}
\put(2,4.1){$E_{2j}$}
\put(8,4.5){\circle*{0.06}}
\put(7.5,4.1){$E_{2j+1}$}

\put(2,4.5){\line(1,0){6}}
\put(5,4.5){\curve(-2.645, 0.951, 0, 0)}
\put(5,4.5){\curve(2.645, 0.951, 0, 0)}
\put(5,4.5){\curve(1, 0, 0.7, 0.7, 0, 1, -0.7, 0.7, -1, 0)}
\curvedashes{0.05,0.05}
\put(5,4.5){\curve(0, 0, -2.645, -0.951)}
\put(5,4.5){\curve(0, 0, 2.645, -0.951)}
\put(5,4.5){\curve(1, 0, 0.7, -0.7, 0, -1, -0.7, -0.7, -1, 0)}
\curvedashes{}


\put(1.4,2.3){$C_{j1}^*$} 
\put(8.0,2.3){$C_{j2}^*$} 
\put(1.4,0.3){$C_{j1}$} 
\put(8.0,0.3){$C_{j2}$} 
\put(5,1.5){\circle*{0.06}}
\put(4.9,1.7){$z_j^*$}
\put(2,1.5){\circle*{0.06}}
\put(2,1.1){$E_{2j}$}
\put(8,1.5){\circle*{0.06}}
\put(7.5,1.1){$E_{2j+1}$}

\put(5,1.5){\curve(0, 0, 2.645, -0.951)}
\put(5,1.5){\curve(0, 0, -2.645, -0.951)}
\put(5,1.5){\curve(1, 0, 0.7, -0.7, 0, -1, -0.7, -0.7, -1, 0)}
\curvedashes{0.05,0.05}
\put(2,1.5){\curve(0,0,6,0)}
\put(5,1.5){\curve(1, 0, 0.7, 0.7, 0, 1, -0.7, 0.7, -1, 0)}
\put(5,1.5){\curve(2.645, 0.951, 0, 0)}
\put(5,1.5){\curve(-2.645, 0.951, 0, 0)}

\end{picture}
\end{center}
\caption{The small cross containing the stationary phase point
$z_j$ and its flipping image containing $z_j^*$. Views from the top 
and bottom sheet. Dotted curves lie in the bottom sheet. \label{fig1}}
\end{figure}

In the previous section we have reduced everything to the solution of the Riemann--Hilbert problem 
\begin{align}\nn
& m^3_+(p) = m^3_-(p) J^3(p),\\ \nn
& (m^3_1) \ge -\dinus{x,t}, \quad (m^3_2) \ge -\dinu{x,t},\\ \nn
& m^3(p^*) = m^3(p) \sigI \\ \nn
& m^3(p_{\infty }) = \begin{pmatrix}  1 & 1 \end{pmatrix},
\end{align}
where the jump matrix $J^3$ is given by \eqref{rhpm5jump}. We have performed a deformation in such a way that the jumps $J^3$ on the oriented paths $C_k$, $C_k^*$ for $k\neq j$ are of the form ``$\id + \text{exponentially small}$'' asymptotically as $t\to \infty$. The same is true for the oriented paths $C_{j1}$, $C_{j2}$, $C_{j1}^*$, $C_{j2}^*$ at least away from the stationary phase points $z_j$, $z^*_j$. The purpose of this section will be to derive the actual asymptotic rate at which $m^3(p)\to \begin{pmatrix}  1 & 1 \end{pmatrix}$ following again \cite{kt2}.
The jump contour near the stationary phase points (cf.~Figure~\ref{fig1}) will be denoted by $\Sigma^C(z_j)$ and $\Sigma^C(z_j^*)$. On these crosses 
the jumps read
\be 
\aligned
J^3&= \ti{B}_+ = \begin{pmatrix}
1 & - \frac{T}{T^*} \frac{R^* \Theta^*}{1-R^* R} \E^{-t\, \phi} \\
0 & 1 \end{pmatrix}, \quad p \in C_{j1}, \\
J^3&= \ti{B}_-^{-1} = \begin{pmatrix} 1 & 0 \\
\frac{T^*}{T}  \frac{R \Theta}{1-R^* R} \E^{t\, \phi} & 1
\end{pmatrix}, \quad p \in C_{j1}^{*}, \\
J^3 &= \ti{b}_+ = \begin{pmatrix} 1 & 0 \\
\frac{T^*}{T} R\Theta \E^{t\, \phi} & 1 \end{pmatrix}, \quad p \in C_{j2}, \\
J^3 &= \ti{b}_-^{-1} = \begin{pmatrix}
1 & -\frac{T}{T^*} R^* \Theta^* \E^{-t\, \phi} \\
0 & 1 \end{pmatrix}, \quad p \in C_{j2}^*.\\
\endaligned
\ee

To reduce our Riemann--Hilbert problem to the one corresponding to the two
crosses we proceed as follows: We take a small disc $D$ around $z_j(x/t)$ and project it
to the complex plane using the canonical projection $\pi$. Now consider the
(holomorphic) Riemann--Hilbert problem in the complex plane with the very jump
obtained by projection and normalize it to be $\id$ near $\infty$. 

The corresponding Riemann--Hilbert problem is solved in \cite[Appendix~A]{krt2}. To apply \cite[Theorem~A.1]{krt2} we need the behavior of the jump matrix $J^3$, that is, the behavior of $T(p,x,t)$ near the stationary phase points $z_j$ and $z_j^*$.

The following lemma gives more information on the singularities of $T(p,x,t)$ near the stationary phase points $z_j$, $j=0,\dots,g$ and the band edges $E_j$, $j=0,\dots,2g+1$ (setting $E_{2g+1}=\infty $).

\begin{lemma} \label{lemd}
For $p$ near a stationary phase point $z_j$ or $z_j^*$ (not equal to a band edge) we have
\be\label{defepm}
T(p,x,t)= (z-z_j)^{\pm\I\nu} e^\pm(z), \quad p=(z,\pm),
\ee
where $e^\pm(z)$ has continuous limits near $z_j$ and
\be \label{defnu}
\nu = - \frac{1}{2\pi} \log (1-|R(z_j)|^2)>0.
\ee
Here $(z-z_j)^{\pm\I\nu}=\exp(\pm\I\nu\log(z-z_j))$, where the branch cut of the logarithm is
along the negative real axis.

For $p$ near a band edge $E_k\in C(x/t)$ we have
\be
T(p,x,t) = T^{\pm 1}(z) \ti{e}^\pm(z), \quad p=(z,\pm),
\ee
where $\ti{e}^\pm(z)$ is holomorphic near $E_k$ if none of the $\nu_j$ is equal
to $E_k$ and $\ti{e}_\pm(z)$ has a first order pole at $E_k=\nu_j$ else.
\end{lemma}

\begin{proof}
By factorizing the jump according to $1-|R(p)|^2=( 1-|R(z_j)|^2)\frac{1-|R(p)|^2}{1-|R(z_j)|^2}$ one can rewrite \eqref{defd}. Then one considers the Abelian differential $\om_{p\, p^*}$ for $p\in \mathcal{K}_g\backslash \{p_{\infty}\}$ which is explicitly given by the formula \eqref{diffpps}.
One has
\be \label{intomppst}
\frac{1}{2}\int_{C(x/t)}\om_{p\, p^*}=\pm \log (z-z_j)\pm \alpha (z_j)+O(z-z_j),\quad p=(z,\pm)
\ee
and thus
\be
\int_{C(x/t)}\om_{p\, p_{\infty}}=\pm \log (z-z_j)\pm \alpha (z_j)+O(z-z_j),\quad p=(z,\pm),
\ee
since $\int_{C(x/t)}f\om_{p\, p_{\infty}}=\frac{1}{2}\int_{C(x/t)}f\om_{p\, p^*}$ for any symmetric function $f(q)=f(q^*)$. From this the first claim follows. For the second claim note that the function
\be \nn
t(p)=\left\{ \begin{array}{ll}
T(z), & p=(z,+)\in \Pi_+, \\ T(z)^{-1}, &p=(z,-)\in \Pi_-,
\end{array}
\right.
\ee
satisfies the following (holomorphic) Riemann--Hilbert problem
\be \nn
\aligned
&t_+(p)=t_-(p)(1-|R(p)|^2),\quad p\in \Sigma,\\ & t(p_{\infty})=1.
\endaligned
\ee
Thus $T(p,x,t)/t(p)$ has no jump along $C(x/t)$ and is therefore holomorphic near $C(x/t)$ away from band edges $E_k=\nu_j$ (where there is a simple pole) by the Schwarz reflection principle.
\end{proof}
Moreover,
\begin{lemma}
We have
\be
e^{\pm}(z) = \ol{e^\mp(z)}, \qquad p=(z,\pm)\in\Sigma\setminus C(x/t)
\ee
and
\be
\aligned
e^+(z_j)=&\exp \big( \I \nu\alpha(z_j)\big)\frac{\theta \big( \ulz(p_{\infty},x,t)+\ul{\delta}(x/t)\big)}{\theta \big(\ulz(p_{\infty},x,t)\big)}\frac{\theta \big(\ulz(z_j,x,t)\big)}{\theta \big(\ulz(z_j,x,t)+ \ul{\delta}(x/t)\big)} \cdot\\
& \cdot \exp\bigg( -\sum_{\rho_k<\zeta(x/t)}\int_{E(\rho_k)}^{\rho_k}\om_{z_j\,z_j^*}+\frac{1}{4\pi\I} \int_{C(x/t)} \log \Big( \frac{1-|R|^2}{1-|R(z_j)|^2}\Big) \om_{z_j\,z_j^*}\bigg),
\endaligned
\ee
where
\be \label{defalphazj}
\alpha(z_j)=\lim_{p\to z_j}\Big( \frac{1}{2}\int_{C(x/t)}\om_{p\,p^*}-\log \big( \pi (p)-z_j\big)\Big).
\ee
Here $\alpha (z_j)\in \R$ and $\om_{p\,p^*}$ is real on $C(x/t)$.
\end{lemma}
\begin{proof}
The first claim follows from the fact that
\be \nn
T(p^*,x,t) = T(\ol{p},x,t)=\ol{T(p,x,t)} \quad \text{for } p\in\Sigma\setminus C(x/t).
\ee
For the second claim just follow the argument used in the proof of the previous lemma.
\end{proof}
By Lemma~\ref{lemd} one deduces that near the stationary phase points the jumps are given by
\be 
\aligned
\hat{B}_+ & = \begin{pmatrix}
1 & - \bigg(\sqrt{\frac{\phi ''(z_j)}{\I }}(z-z_j) \bigg)^{2\I\nu} \frac{\ol{r}}{1-|r|^2} \E^{-t\, \phi} \\
0 & 1 \end{pmatrix}, \quad p \in L_{j1}, \\
\hat{B}_-^{-1} & = \begin{pmatrix} 1 & 0 \\
\bigg(\sqrt{\frac{\phi ''(z_j)}{\I }}(z-z_j) \bigg)^{-2\I\nu} \frac{r}{1-|r|^2} \E^{t\, \phi} & 1
\end{pmatrix}, \quad p \in L_{j1}^{*}, \\
\hat{b}_+ & = \begin{pmatrix} 1 & 0 \\
\bigg(\sqrt{\frac{\phi ''(z_j)}{\I }}(z-z_j) \bigg)^{-2\I\nu} r \E^{t\, \phi} & 1 \end{pmatrix}, \quad p \in L_{j2}, \\
\hat{b}_-^{-1} &= \begin{pmatrix}
1 & -\bigg(\sqrt{\frac{\phi ''(z_j)}{\I }}(z-z_j) \bigg)^{2\I\nu} \ol{r} \E^{-t\, \phi} \\
0 & 1 \end{pmatrix}, \quad p \in L_{j2}^*,\\
\endaligned
\ee
where (cf.~eq.~\eqref{defepm})
\be\label{defr}
r=R(z_j) \Theta(z_j,x,t) \frac{\ol{e^+(z_j)}}{e^+(z_j)}\Big( \frac{\phi ''(z_j)}{\I }\Big) ^{\I \nu}.
\ee
The error terms will satisfy appropriate H\"older estimates, that is
\be
\| \hat{B}_+(p) - \hat{B}_+(p)\| \le C |z-z_j|^\alpha, \qquad p=(z,+) \in C_{j1},
\ee
for any $\alpha<1$ and similarly for the other matrices. Thus the assumptions of \cite[Theorem~A.1]{krt2} are satisfied and we can conclude that the solution on $\pi \big( \Sigma^C(z_j)\big)$ is of the form
\be
M(z)  =  \id + \frac{M_0}{z-z_j} \frac{1}{t^{1/2}} + O(t^{-\alpha}),
\ee
where
\begin{align}
M_0 &= \I \sqrt{\I/\phi''(z_j)} \begin{pmatrix} 0 & -\beta(t)\\ \ol{\beta(t)} & 0 \end{pmatrix},\\
\beta(t) &= \sqrt{\nu} \E^{\I(\pi/4-\arg(r)+\arg(\Gamma(\I\nu)))} \E^{-t \phi(z_j)} t^{-\I\nu},
\end{align}
and $1/2<\alpha <1$.
Now we lift this solution in the complex plane back to the small disc $D$ on the Riemann surface $\mathcal{K}_g$ by setting
\be
M(p)=\left\{ \begin{array}{ll}
M(z), & p\in D, \\
\sigI M(z) \sigI, & p\in D^*.
\end{array} \right.
\ee
Thus we conclude that the solution on $\Sigma ^{C}(z_j)$ is given by
\be
M^C(p)=\id +\frac{1}{t^{1/2}}\frac{M_0}{z-z_j}+O(t^{-\alpha}), \quad p=(z,+),
\ee
and the one on $\Sigma ^{C}(z_j^*)$ reads
\be
\ti{M}^C(p)=\id +\frac{1}{t^{1/2}}\frac{\overline{M_0}}{z-z_j}+O(t^{-\alpha}), \quad p=(z,-).
\ee
Then
\be
m^4(p) = \begin{cases}
m^3(p) M^{C}(p)^{-1}, & p\in D,\\
m^3(p) \ti{M}^C(p)^{-1}, & p\in D^*,\\
m^3(p), & \text{else},
\end{cases}
\ee
has no jump inside $D\cup D^*$ but jumps on the boundary $\partial D \cup \partial D^*$. All jumps outside $D \cup D^*$ are of the form $\id+ exponentially~small$ and the jump on $\partial D \cup \partial D^*$ is of the form $\id + O(t^{-1/2})$. 
In order to identify the leading behavior it remains to rewrite the Riemann--Hilbert problem for $m^4$ as a singular integral equation following Section~\ref{sec:sieq}.
Let the operator $C_{w^4}: L^2(\Sigma^4) \to L^2(\Sigma^4) $ be defined by
\be
C_{w^4} f = C_- (f w^4)
\ee
for a vector valued $f$, where $w^4 = J^4 - \id$
and
\be
(C_\pm f)(q) = \lim_{p \to q \in \Sigma^4} \frac{1}{2 \pi \I} \int_{\Sigma^4} f\, \ul{\Omega}_p^\uhnuz, \qquad
\ul{\Omega}_p^\uhnuz = \begin{pmatrix}
\Omega_p^{\uhnuz^*,p_{\infty}} & 0 \\ 0 & \Omega_p^{\uhnuz,p_{\infty}}
\end{pmatrix},
\ee
are the Cauchy operators for our Riemann surface. In particular, $\Omega_p^{\uhnuz,q}$
is the Cauchy kernel given by
\be\label{defOm}
\Omega_p^{\uhnuz,q} = \om_{p\, q} + \sum_{j=1}^g I_j^{\uhnuz,q}(p) \zeta_j,
\ee
where
\be
I_j^{\uhnuz,q}(p) = \sum_{\ell=1}^g c_{j\ell}(\uhnuz) \int_q^p \om_{\hnu_\ell,0}.
\ee
Here $\om_{q,0}$ is the (normalized) Abelian differential of the second kind with a second order pole at $q$ (cf.\ Remark~\ref{remabdiff2k} below) and $\om_{p\, q}$ denotes the Abelian differential of the third kind with simple poles at $p$ and $q$.
Note that $I_j^{\uhnuz,q}(p)$ has first order poles at the points $\uhnuz$.

The constants $c_{j\ell}(\uhnuz)$ are chosen such that $\Omega_p^{\uhnuz,q}$ is single
valued, that is,
\be\label{defcjlnu}
\left( c_{\ell k}(\uhnuz) \right)_{1 \le \ell,k \le g} = 
\left( \sum_{j=1}^g c_k(j) \frac{\mu_\ell^{j-1}d\pi}{R_{2g+2}^{1/2}(\hmu_\ell)} \right)_{1 \le \ell,k \le g}^{-1}
\ee
where $c_k(j)$ are defined in \eqref{defcjk} (cf.\ Lemma~\ref{lemck}).

Consider the solution $\mu^4$ of the singular integral equation
\be 
\mu = \rI + C_{w^4} \mu  \quad\text{ in }\quad L^2(\Sigma^4).
\ee
Then the solution of our Riemann--Hilbert problem is given by
\be
m^4(p) = 
\rI + \frac{1}{2\pi\I} \int_{\Sigma^4} \mu^4 \, w^4 \, \ul{\Omega}_{p}^\uhnuz.
\ee
By $\|w^4\|_\infty = O(t^{-1/2})$ Neumann's formula implies
\be
\mu^4(q) = (\id - C_{w^4})^{-1} \rI = \rI + O(t^{-1/2}).
\ee
Moreover,
\be
w^4(p) = \begin{cases} -\frac{M_0}{z-z_j} \frac{1}{t^{1/2}} + O(t^{-\alpha}), & p \in \partial D,\\
-\frac{\ol{M_0}}{z-z_j} \frac{1}{t^{1/2}} + O(t^{-\alpha}), & p \in \partial D^*.
\end{cases}
\ee
Hence we obtain
\begin{align}\nn \label{asym4}
m^4(p)= & \rI - \frac{\rI M_0}{t^{1/2}} \frac{1}{2\pi\I} \int_{\partial D} \frac{1}{\pi-z_j} \, \ul{\Omega}_{p}^\uhnuz\\ \nn
& - \frac{\rI \ol{M_0}}{t^{1/2}} \frac{1}{2\pi\I} \int_{\partial D^*} \frac{1}{\pi-z_j} \, \ul{\Omega}_{p}^\uhnuz
+ O(t^{-\alpha})\\ \nn
= & \rI - \frac{\rI M_0}{t^{1/2}} \ul{\Omega}_{p}^\uhnuz(z_j)
- \frac{\rI \ol{M_0}}{t^{1/2}} \ul{\Omega}_{p}^\uhnuz(z_j^*)+ O(t^{-\alpha})\\ \nn
= & \rI - \sqrt{\frac{\I}{\phi''(z_j)t}} \times \nn \\
&\times \begin{pmatrix} \I \ol{\beta} \Omega_p^{\uhnuz^*,p_{\infty}}(z_j) - \I \beta\Omega_p^{\uhnuz^*,p_{\infty}}(z_j^*) &
- \I \beta \Omega_p^{\uhnuz,p_{\infty}}(z_j) + \I \ol{\beta} \Omega_p^{\uhnuz,p_{\infty}}(z_j^*) \end{pmatrix}  \nn\\ &+O(t^{-\alpha}).
\end{align}
Since we need the asymptotic expansions around $p_{\infty}$ we note

\begin{lemma}
We have 
\be \label{expOmega}
\Omega_p^{\uhnuz,p_{\infty}}(z_j)= \Lambda^\uhnuz_1(z_j) \zeta +\Lambda^\uhnuz_2(z_j) \zeta^2+ O\big(\zeta^3\big)
\ee
for $\zeta=z^{-1/2}$ being the local chart near $p_{\infty}$ and
\begin{align}
\Lambda^\uhnuz_1(z_j) &= \om_{p_{\infty},0}(z_j) - \sum_{k=1}^g\sum_{\ell =1}^g c_{k\ell}(\uhnuz) \alpha_{g-1}(\hat\nu_{\ell}) \zeta_k(z_j),\\
\Lambda^\uhnuz_2(z_j) &= \om_{p_{\infty},1}(z_j) - \frac{1}{2}\sum_{k=1}^g\sum_{\ell =1}^g c_{k\ell}(\uhnuz) \zeta_k(z_j),
\end{align}
where $\om_{q,k}$, $k=0,1,\dots$, is an Abelian differential of the second kind with a single pole of order $k+2$ at $q$ and $\alpha_{g-1}(\hat\nu_{\ell})$ denotes a constant defined in Remark~\ref{remabdiff2k} below.
\end{lemma}

\begin{proof}
Use the local coordinate $\zeta =z^{-1/2}$ near $p_{\infty}=(\infty ,\infty )$ and expand the differential $\om_{pp_{\infty}}$ like it is done in \cite[Theorem 4.1]{tag} and $\int_{p_{\infty}}^{p} \om_{\hat\nu_\ell ,0}$ by using the expression \eqref{dsknu}. For $g\geq1$ one gets
\[
\om_{\hat\nu,0}(\zeta)=-\alpha_{g-1}(\hat\nu)-\zeta+O(\zeta^2)
\]
and thus the claimed formulas for $\Lambda^\uhnuz_1(z_j)$ and $\Lambda^\uhnuz_2(z_j)$ follow.
\end{proof}

\begin{remark}\label{remabdiff2k}
The Abelian differential appearing in the previous lemma is explicitly given by
\be \label{dsknu}
\om_{\hat\nu,0} =  \frac{\Rgo  + \Rg{\hat\nu} +
\frac{R_{2g+1}'(\hat\nu)}{2\Rg{\hat\nu}} (\pi-\nu) +
P_{\hat\nu,0} \cdot (\pi-\nu)^2 }{2 (\pi -\nu)^2\Rgo}d\pi,
\ee
with $P_{\hat\nu,0}$ a polynomial of degree $g-1$ which has to be determined from the
normalization. We will use the notation
\be\label{equ:diff2kal}
P_{\hat\nu,0}(z)=\sum_{j=0}^{g-1}\alpha_j(\hat\nu)z^j.
\ee
Concerning the Abelian differential $\om_{p_{\infty},0}$ we refer to \eqref{dsk0}. The differential $\om_{p_{\infty},1}$ is given by
\be \label{dsk1}
\omega_{p_\infty,1} = \Big( - \frac{\Rgo }{2} + P_{p_\infty,1} \Big) \frac{d\pi}{\Rgo },
\ee
where $P_{p_\infty,1}$ is a polynomial of degree $g-1$ which has to be determined by the vanishing $a_j$-periods as usual.
\end{remark}
Note that the following relations are valid
\be \label{equ:propdiff}
\aligned
\om_{p_{\infty},0}(z_j^*)&=-\om_{p_{\infty},0}(z_j),\\
\om_{p_{\infty},1}(z_j^*)&+\om_{p_{\infty},1}(z_j)=-1,
\endaligned
\ee
and 
\be \label{equ:propcoeff}
c_{k\ell}(\uhnuz^*)=-c_{k\ell}(\uhnuz), \qquad \zeta_k(z_j^*)=-\zeta_k(z_j).
\ee
Moreover, the coefficients $\alpha_j(\hat\nu)$, $j=0,\dots,g-1$ of the polynomial $P_{\hat\nu,0}$ fulfill the relation
\be \label{equ:coeffalpha}
\alpha_j(\hat{\nu}^*)=-\alpha_j(\hat{\nu}),\qquad j=0,\dots,g-1.
\ee

Now we come to prove Theorem~\ref{thmMain3}. As in the previous section, the asymptotics can be read off by using
\be \label{relmm4}
m(p) =  m^4(p) \begin{pmatrix} \frac{1}{T(p^*,x,t)} & 0 \\ 0 & \frac{1}{T(p,x,t)} \end{pmatrix}
\ee
for $p$ near $p_{\infty}$ and comparing with \eqref{m2infp}. 
From that one deduces
\begin{align*}
m_2(p)=& m_2^4(p)T(p)^{-1}= \\
=&1+\Big( \sqrt{\frac{\I }{\phi ''(z_j)t}}\big( \I\beta \Lambda^{\uhnuz}_1(z_j)-\I\ol{\beta} \Lambda^{\uhnuz}_1(z_j^*)\big)-T_1(x,t)+O(t^{-\alpha})\Big) \frac{1}{\sqrt{z}}+O( z^{-1}) ,
\end{align*}
where we have used \eqref{asym4}, \eqref{expOmega} and \eqref{Tinfp}. Comparing this asymptotic expansion with \eqref{m2infp} yields
\[
\int_x^{+\infty}(V-V_q)(y)dy=2\sqrt{\frac{\I}{\phi''(z_j)t}}\big(\beta\Lambda_1^{\uhnuz}(z_j)-\ol{\beta}\Lambda_1^{\uhnuz}(z_j^*)\big)+2\I T_1(x,t)+O(t^{-\alpha}).
\]
Invoking \eqref{equ:propdiff}, \eqref{equ:propcoeff} and \eqref{equ:coeffalpha} one gets
\be \label{equ:Lambda1}
\aligned
\Lambda_1^{\uhnuz}(z_j^*)&=-\Lambda_1^{\uhnuz}(z_j),\\
\Lambda_1^{\uhnuz^*}(z_j)&=\Lambda_1^{\uhnuz}(z_j),
\endaligned
\ee
and therefore
\be
\int_x^{\infty}(V-V_q)(y,t)dy=4\sqrt{\frac{\I }{\phi ''(z_j)t}}\re \big(\beta (x,t)\big) \Lambda^{\uhnuz}_1(x,t)+2\I T_1(x,t)+O(t^{-\alpha}).
\ee
Finally, using the definition of the limiting solution \eqref{limlatsol} proves the claim. Note that one obtains the same result if one compares the expressions for the component $m_1$.

Similarly one obtains \eqref{eqvvl} by using \eqref{equ:expm1m2} instead of \eqref{m2infp}.

\section{Analytic Approximation}
\label{secanap}

In this section we want to show how to get rid of the analyticity assumption on the reflection coefficient $R(p)$.
To this end we will split $R(p)$ into an analytic part $R_{a,t}$ plus a {\em small} rest $R_{r,t}$ following
the ideas of \cite{dz} (see also \cite[Sect.~6]{krt2}). The analytic part will be moved to regions of the Riemann surface while the rest
remains on $\Sigma=\pi^{-1}\big( \sigma(H_q)\big)$. This needs to be done in such a way that the rest
is of $O(t^{-1})$ and the growth of the analytic part can be controlled by the decay of the phase.

In order to avoid problems when one of the poles $\nu_j$ hits $\Sigma$, we have to make the
approximation in such a way that the nonanalytic rest vanishes at the band edges. That is, split $R$ according to
\begin{align} \nn
R(p) =& R(E_{2j}) \frac{(z-E_{2j})(E_{2j+1}-\I)}{(E_{2j+1}-E_{2j})(z-\I)} + R(E_{2j+1}) \frac{(z-E_{2j+1})(E_{2j}-\I)}{(E_{2j}-E_{2j+1})(z-\I)}\\ \label{splitR}
& + Q_j(p) \ti{R}(p), \qquad p=(z,\pm),
\end{align}
where $Q_j(p)$ is a rational function with first order zeros at $E_{2j}, E_{2j+1}$ and with all other zeros and poles away from $\Sigma$,
and approximate $\ti{R}$. Note that if $R \in C^l(\Sigma)$, then $\ti{R} \in C^{l-1}(\Sigma)$.

We will use different splittings for different bands depending on whether the band contains our stationary phase
point $z_j(x/t)$ or not. We will begin with some preparatory lemmas.

For the bands containing no stationary phase points we will use a splitting based on the following Fourier transform 
associated with the background operator $H_q$. Given $R \in C^l(\Sigma)$ we can write
\be
R(p) =\int_\R \hat{R}(x)\psi_q(p,x,0)dx,
\ee
where $\psi_q(p,x,t)$ denotes the time-dependent Baker--Akhiezer function and (cf.\ \cite{egt}, \cite{et})
\be
\hat{R}(x) = \frac{1}{2\pi\I} \oint_\Sigma R(p) \psi_q(p,x,0)  \frac{\I \prod_{j=1}^g (\pi(p)-\mu_j)}{2\Rg{p}}d\pi(p).
\ee
If we make use of \eqref{defpsiq}, the above expression for $R(p)$ is of the form
\be
R(p)= \int_\R \hat{R}(x) \theta_q(p,x,0)\exp \big( \I x k(p) \big) dx.
\ee
where $k(p)= -\I \int_{E_0}^p \omega_{p_{\infty},0}$ and $\theta_q(p,n,t)$ collects the remaining
parts in \eqref{defpsiq}.

Using $k(p)$ as a new coordinate and performing $l$ integration by parts one obtains (cf.\ \cite{et})
\be
|\hat{R}(x)| \le \frac{const}{1+|x|^l}
\ee
provided $R \in C^l(\Sigma)$.

\begin{lemma}\label{lem:analapprox}
Suppose $\hat{R} \in L^1(\R)$, $x^l \hat{R}(x) \in L^1(\R)$ and let $\beta>0$ be given.
Then we can split $R(p)$ according to
\[
R(p)= R_{a,t}(p) + R_{r,t}(p),
\]
such that $R_{a,t}(p)$ is analytic for in the region $0 < \im(k(p)) <\eps$ and 
\begin{align}
|R_{a,t}(p) \E^{-\beta t} | &= O(t^{-l}), \quad 0 < \im(k(p)) <\eps,\\
|R_{r,t}(p)| &= O(t^{-l}), \quad p\in\Sigma.
\end{align}
\end{lemma}

\begin{proof}
We choose
\[
R_{a,t}(p) = \int_{x=-K(t)}^\infty \hat{R}(x) \theta_q(p,x,0)\exp \big( \I x k(p) \big) dx
\]
with $K(t) = \frac{\beta_0}{\eps} t$ for some positive $\beta_0<\beta$. Then, for 
$0< \im(k(p)) <\eps$,
\begin{align*}
\left\vert R_{a,t}(k)\E^{-\beta t} \right\vert 
& \leq C \E^{-\beta t} \int_{x=-K(t)}^\infty | \hat{R}(x) | \E^{-\im(k(p)) x} dx\\
& \leq C \E^{-\beta t}\E^{K(t)\eps}\| F \|_1
= \|\hat{R}\|_1 \E^{-(\beta-\beta_0)t},
\end{align*}
which proves the first claim.
Similarly, for $p\in\Sigma$,
\[
\vert R_{r,t}(k) \vert 
\leq C \int_{x=K(t)}^{\infty} \frac{x^l |\hat{R}(-x)|}{x^l} dx
\leq C \frac{\|x^l \hat{R}(-x)\|_1}{K(t)^l} 
\leq \frac{\tilde{C}}{t^{l}}  
\]
\end{proof}

For the band which contains $z_j(x/t)$ we need to take the small vicinities of the stationary phase points into account.
Since the phase is cubic near these points, we cannot use it to dominate the exponential growth of the analytic
part away from $\Sigma$. Hence we will take the phase as a new variable and use the Fourier transform
with respect to this new variable. Since this change of coordinates is singular near the stationary phase points,
there is a price we have to pay, namely, requiring additional smoothness for $R(p)$. 

Without loss of generality we will choose the path of integration in our phase $\phi(p)$, defined in \eqref{defsp},
such that $\phi(p)$ is continuous (and thus analytic) in $D_{j,1}$ with continuous limits on the boundary
(cf.\ Figure~\ref{fig2}). We begin with

\begin{lemma}
Suppose $R(p)\in C^5(\Sigma)$. Then we can split $R(p)$ according to
\be
R(p) = R_0(p) +\frac{z- z_j}{z-\I} H(p), \qquad p=(z,\pm) \in \Sigma \cap D_{j,1},
\ee
where $R_0(p)$ is a real rational function on $\M$ such that $H(p)$ vanishes  
at $z_j$, $z_j^*$ of order three and has a Fourier transform
\be
H(p)=\int_\R \hat{H}(x) \E^{x \phi(p)} dx,
\ee
with $x\hat{H}(x)$ integrable. Here $\phi$ denotes the phase defined in \eqref{defsp}.
\end{lemma}

\begin{proof}
We begin by choosing a rational function $R_0(p) = a(z) + b(z) \Rg{p}$ with $p=(z,\pm)$ such that $a(z)$, $b(z)$
are real-valued rational functions which are chosen such that $a(z)$ matches the values of $\re(R(p))$
and its first four derivatives at $z_j$ and $\I^{-1} b(z) \Rg{p}$ matches the values of $\im(R(p))$
and its first four derivatives at $z_j$. Moreover, all poles are chosen away from $\Sigma$.
Since $R(p)$ is $C^5$ we infer that $H(p)\in C^4(\Sigma)$
and it vanishes together with its first three derivatives at $z_j$, $z_j^*$.

Note that $\phi(p)/\I$, where $\phi$ is defined in \eqref{defsp} has a maximum at $z_j^*$
and a minimum at $z_j$. Thus the phase $\phi(p)/\I$ restricted to $\Sigma \cap D_{j,1}$ gives
a one to one coordinate transform $\Sigma \cap D_{j,1} \to [\phi(z_j^*)/\I, \phi(z_j)/\I]$   
and we can hence express $H(p)$ in this new coordinate (setting it equal to zero outside this interval). The coordinate  
transform locally looks like a cube root near $z_j$ and $z_j^*$,
however, due to our assumption that $H$ vanishes there, $H$ is still  
$C^2$ in this new coordinate and the Fourier transform
with respect to this new coordinates exists and has the required  
properties.
\end{proof}

Moreover, as in Lemma~\ref{lem:analapprox} we obtain:

\begin{lemma}\label{lem:analapprox2}
Let $H(p)$ be as in the previous lemma. Then we can split $H(p)$ according to
$H(p)= H_{a,t}(p) + H_{r,t}(p)$ such that $H_{a,t}(p)$ is analytic in the region $\re(\phi(p))<0$
and 
\be
|H_{a,t}(p) \E^{\phi(p) t/2} | = O(1), \: p\in \ol{D_{j,1}}, \quad
|H_{r,t}(p)| = O(t^{-1}), \: p\in\Sigma.
\ee
\end{lemma}

\begin{proof}
We choose $H_{a,t}(p) = \int_{x=-K(t)}^\infty \hat{H}(x)\E^{x \phi(p)} dx$ with $K(t) = t/2$.
Then we can proceed as in Lemma~\ref{lem:analapprox}:
\begin{align*}
\vert H_{a,t}(p) \E^{\phi(p) t/2} \vert
\leq \|\hat{H}\|_1 |\E^{-K(t) \phi(p)+\phi(p) t/2}|
\leq \|\hat{H}\|_1
\end{align*} 
and 
\[
|H_{r,t}(p)| \leq \frac{1}{K(t)} \int_{x=K(t)}^\infty x |\hat{H}(-x)| dx \leq \frac{C}{t}.
\]
\end{proof}

Clearly an analogous splitting exists for $p\in\Sigma \cap D_{j2}$.

Now we are ready for our analytic approximation step. First of all recall that our jump is given in terms
$\ti{b}_\pm$ and $\ti{B}_\pm$ defined in \eqref{defbt} and \eqref{defBt}, respectively. While $\ti{b}_\pm$
are already in the correct form for our purpose, this is not true for $\ti{B}_\pm$ since they contain
the non-analytic expression $|T(p)|^2$. To remedy this we will rewrite $\ti{B}_\pm$ in terms of the left
rather than the right scattering data. For this purpose let us use the notation $R_r(p) \equiv R_+(p)$
for the right and $R_l(p) \equiv R_-(p)$ for the left reflection coefficient. Moreover, let $T_r(p,x,t) \equiv T(p,x,t)$ be the right and
$T_l(p,x,t) \equiv T(p)/T_r(p,x,t)$ be the left partial transmission coefficient.

With this notation we have
\be
J^2(p) = \begin{cases}
\ti{b}_-(p)^{-1} \ti{b}_+(p), \qquad \pi(p)> \zeta(x/t),\\
\ti{B}_-(p)^{-1} \ti{B}_+(p), \qquad \pi(p)< \zeta(x/t),\\
\end{cases}
\ee
where
\begin{align*}
\ti{b}_- &= \begin{pmatrix} 1 & \frac{T_r(p,x,t)}{T_r(p^*,x,t)}R_r(p^*)\Theta(p^*)\E^{-t\phi(p)} \\ 0 & 1 \end{pmatrix}, \\
\ti{b}_+ &= \begin{pmatrix} 1 & 0 \\ \frac{T_r(p^*,x,t)}{T_r(p,x,t)}R_r(p)\Theta(p)\E^{-t\phi(p)}& 1 \end{pmatrix},
\end{align*}
and
\begin{align*}
\ti{B}_- &= \begin{pmatrix} 1 & 0 \\ -\frac{T_{r,-}(p^*,x,t)}{T_{r,-}(p,x,t)}  \frac{R_r(p)  \Theta(p)}{|T(p)|^2} \E^{t\, \phi(p)} & 1 \end{pmatrix}, \\
\ti{B}_+ &= \begin{pmatrix} 1 &  - \frac{T_{r,+}(p,x,t)}{T_{r,+}(p^*,x,t)} \frac{R_r(p^*) \Theta(p^*)}{|T(p)|^2} \E^{-t\, \phi(p)} \\ 0 & 1 \end{pmatrix}.
\end{align*}
Using \eqref{reltrpm} we can write
\begin{align*}
\ti{B}_- &= \begin{pmatrix} 1 & 0 \\ \frac{T_l(p^*,x,t)}{T_l(p,x,t)}R_l(p)\Theta(p)\E^{-t\phi(p)} & 1 \end{pmatrix}, \\
\ti{B}_+ &= \begin{pmatrix} 1 & \frac{T_l(p,x,t)}{T_l(p^*,x,t)}R_l(p^*)\Theta(p^*)\E^{-t\phi(p)} \\ 0 & 1 \end{pmatrix}.
\end{align*}
Now we split $R_r(p)=R_{a,t}(p)+R_{r,t}(p)$ by splitting $\ti{R}_r(p)$ defined via \eqref{splitR} according to Lemma~\ref{lem:analapprox}
for $\pi(p)\in [E_{2k},E_{2k+1}]$ with $k< j$ (i.e., not containing $\zeta(x/t)$) and according to Lemma~\ref{lem:analapprox2}
for $\pi(p)\in [E_{2j},\zeta(x/t)]$. In the same way we split $R_l(p)=R_{a,t}(p)+R_{r,t}(p)$ for $\pi(p)\in [\zeta(x/t),E_{2j+1}]$
and $\pi(p)\in [E_{2k},E_{2k+1}]$ with $k> j$. For $\beta$ in Lemma~\ref{lem:analapprox} we can choose
\be
\beta=\left\{ \begin{array}{ll} \min_{p\in C_k} -\re(\phi(p))>0, & \pi(p)>\zeta(x/t),\\
\min_{p\in C_k} \re(\phi(p))>0, & \pi(p)<\zeta(x/t). \end{array}\right.
\ee

In this way we obtain
\begin{align*}
\ti{b}_\pm(p) &= \ti{b}_{a,t,\pm}(p) \ti{b}_{r,t,\pm}(p) = \ti{b}_{r,t,\pm}(p) \ti{b}_{a,t,\pm}(p),\\
\ti{B}_\pm(p) &= \ti{B}_{a,t,\pm}(p) \ti{B}_{r,t,\pm}(p) = \ti{B}_{r,t,\pm}(p) \ti{B}_{a,t,\pm}(p).
\end{align*}
Here $\ti{b}_{a,t,\pm}(p)$, $\ti{b}_{r,t,\pm}(p)$ denote the matrices obtained from $\ti{b}_\pm(p)$ by replacing $R_r(p)$ with
$R_{a,t}(p)$, $R_{r,t}(p)$, respectively. Similarly for $\ti{B}_{a,t,\pm}(p)$, $\ti{B}_{r,t,\pm}(p)$.
Now we can move the analytic parts into regions of the Riemann surface as in Section~\ref{secSPP}
while leaving the rest on $\Sigma$. Hence, rather than \eqref{rhpm5jump}, the jump now reads
\be
J^3(p) = \left\{ \begin{array}{ll}
\ti{b}_{a,t,+}(p), & p\in C_k, \quad \pi(p)>\zeta(x/t), \\
\ti{b}_{a,t,-}(p)^{-1}, &p\in C_k^*, \quad \pi(p)>\zeta(x/t),\\
\ti{b}_{r,t,-}(p)^{-1} \ti{b}_{r,t,+}(p), & p\in \pi^{-1}((\zeta(x/t),+\infty)), \\
\ti{B}_{a,t,+}(p), & p\in C_k, \quad \pi(p)<\zeta(x/t), \\
\ti{B}_{a,t,-}(p)^{-1}, &p\in C_k^*, \quad \pi(p)<\zeta(x/t),\\
\ti{B}_{r,t,-}(p)^{-1} \ti{B}_{r,t,+}(p), & p\in \pi^{-1}(-\infty,\zeta(x/t))). 
\end{array} \right.
\ee
By construction $R_{a,t}(p) = R_0(p) + (\pi(p)- \pi(z_j)) H_{a,t}(p)$ will satisfy the required
Lipschitz estimate in a vicinity of the stationary phase points (uniformly in $t$) and the
jump will be $J^3(p) = \id+O(t^{-1})$. The remaining parts of $\Sigma$ can be handled analogously
and hence we can proceed as in Section~\ref{secCROSS}.

\appendix

\section{Singular integral equations}
\label{sec:sieq}

In the complex plane, the solution of a Riemann--Hilbert problem
can be reduced to the solution of a singular integral equation (see \cite{bc}).
In our case the underlying space is a Riemann surface $\mathbb{M}$.
The purpose of this appendix is to generalize this approach to
meromorphic vector Riemann--Hilbert problem with simple poles at $\rho$, $\rho^{\ast}$
of the type
\begin{align}\nn
& m_+(p) = m_-(p) J(p), \qquad p\in \Sigma,\\ \nn
& (m_1) \ge -\dimuzs - \dirho, \qquad (m_2) \ge -\dimuz - \dirhos,\\ \label{rhpm}
& \Big( m_1(p) - \frac{R^{1/2}_{2g+2}(\rho)}{\prod_{k=1}^g (\rho-\mu_k)}
\frac{\gam_j}{\pi(p)-\rho} \frac{\psi_q(p)}{\psi_q(p^*)} m_2(p) \Big) \ge - \dimuz^*,
\mbox{ near $\rho$},\\ \nn
& \Big( - \frac{R^{1/2}_{2g+2}(\rho)}{\prod_{k=1}^g (\rho-\mu_k)}
\frac{\gam}{\pi(p)-\rho} \frac{\psi_q(p)}{\psi_q(p^*)}  m_1(p) + m_2(p) \Big) \ge - \dimuz,
\mbox{ near $\rho^*$},\\ \nn
& m(p^*) = m(p) \sigI,\\ \nn
& m(p_\infty) = \begin{pmatrix} 1 & 1\end{pmatrix},
\end{align}
Since we require the symmetry condition \eqref{eq:symcond} for our Riemann--Hilbert
problems, we need to adapt the usual Cauchy kernel to preserve this symmetry.
Moreover, we keep the single soliton as an inhomogeneous term which will play
the role of the leading asymptotics in our applications.

Concerning the jump contour $\Sigma$ and the jump matrix $J$ we will
make the following assumptions:

\begin{hypothesis}\label{hyp:rhp}
Let $\Sigma$ consist of a finite number of smooth oriented finite curves in $\mathbb{M}$
which intersect at most finitely many times with all intersections being transversal.
Assume that the contour $\Sigma$ does not contain any of the points $\uhmuz$ and is invariant under
$p\mapsto p^{\ast}$. It is oriented such that under the mapping $p\mapsto p^{\ast}$
sequences converging from the positive sided to $\Sigma$ are mapped to sequences
converging to the negative side. The divisor $\dimuz$ is nonspecial.

The jump matrix $J$ is nonsingular and can be factorized according to
$J = b_-^{-1} b_+ = (\id-w_-)^{-1}(\id+w_+)$, where $w_\pm = \pm(b_\pm-\id)$ are
H\"older continuous and satisfy
\be\label{eq:wpmsym}
w_\pm(p^\ast) = -\sigI w_\mp(p) \sigI,\quad p\in\Sigma.
\ee
Moreover,
\be\label{hyp:normw}
\|w\|_\infty = \|w_+\|_{L^\infty(\Sigma)} + \|w_-\|_{L^\infty(\Sigma)}<\infty.
\ee
\end{hypothesis}

\begin{remark}
The assumption that that none of the poles $\uhmuz$ lie on our contour $\Sigma$ can
be made without loss of generality if the jump is analytic since we can move the contour
a little without changing the value at $p_{\infty}$ (which is the only value we are eventually interested in).
Alternatively, the case where one (or more) of the poles $\hat\mu_j$ lies on $\Sigma$ can be included if one
assumes that $w_\pm$ has a first order zero at $\hat\mu_j$. In fact, in this case one can replace
$\mu(s)$ by $\ti{\mu}(s)=(\pi(s)-\mu_j)\mu(s)$ and $w_\pm(s)$ by $\ti{w}_\pm(s)=(\pi(s)-\mu_j)^{-1}w_\pm(s)$.

Otherwise one could also assume that the matrices $w_\pm$ are H\"older continuous and
vanish at such points. Then one can work with the weighted measure $-\I\Rg{p}d\pi$ on $\Sigma$.
In fact, one can show that the Cauchy operators are still bounded in this weighted Hilbert space
(cf.\ \cite[Thm.~4.1]{gk}).
\end{remark}

Our first step is to replace the classical Cauchy kernel by  a "generalized" Cauchy
kernel  appropriate to our Riemann surface.
In order to get a single valued kernel we need again
to admit $g$ poles. We follow the construction from \cite[Sec.\ 4]{ro}/

\begin{lemma}[\cite{kt2,krt2}]\label{lemck}
Let $\dimuz$ be nonspecial and introduce the differential
\be\label{defOmpmu}
\Omega_p^{\uhmuz,\rho} = \om_{p\, \rho} + \sum_{j=1}^g I_j^{\uhmuz,\rho}(p) \zeta_j,
\ee
where
\be
I_j^{\uhmuz,\rho}(p) = \sum_{\ell=1}^g c_{j\ell}(\uhmuz) \int_\rho^p \om_{\hmu_\ell,0}.
\ee
Here $\om_{q,0}$ is the (normalized) Abelian differential of the second kind with
a second order pole at $q$ (cf.\ Remark~\ref{remabdiff2k}) and the matrix $c_{j\ell}$
is defined as the inverse matrix of $\eta_\ell(\hmu_j)$, where
$\zeta_\ell = \eta_\ell(z) dz$ is the chart expression in a local chart near $\hmu_j$
(the same chart used to define $\om_{\hmu_j,0}$).

Then $\Omega_p^{\uhmuz,\rho}$ is single valued as a function of $p$ with
first order poles at the points $\uhmuz$.
\end{lemma}

Next we show that the Cauchy kernel introduced in \eqref{defOmpmu} has indeed the correct properties.
We will abbreviate $L^p(\Sigma)=L^p(\Sigma,\C^2)$.

\begin{theorem}[\cite{kt2,krt2}] \label{thmCrs}
Set
\be
\ul{\Omega}_p^{\uhmuz,\rho} = \begin{pmatrix} \Omega_{p}^{\uhmuz^*,\rho^*} & 0 \\                                       
0 & \Omega_p^{\uhmuz,\rho} \end{pmatrix},
\ee
and define the matrix operators as follows. Given a $2\times2$ matrix $f$ defined 
on $\Sigma$ with H\"older continuous entries, let
\be
(C f)(p) = \frac{1}{2\pi\I} \int_\Sigma f(s) \ul{\Omega}_p^{\uhmuz,\rho},
\quad\text{for}\quad p \not \in \Sigma,
\ee
and
\be \label{defCpm}
(C_\pm f)(q) = \lim_{p \to q \in \Sigma} (C f)(p)
\ee
from the left and right of $\Sigma$ respectively (with respect to its orientation).
Then 
\begin{enumerate}
\item
The operators $C_\pm$ are given by the Plemelj formulas
\[
\aligned
(C_+ f)(q) - (C_- f)(q) &= f(q),\\
(C_+ f)(q) + (C_- f)(q) &= \frac{1}{\pi \I}\; \dashint_\Sigma f\, \ul{\Omega}_q^{\uhmuz,\rho},
\endaligned
\]
and extend to bounded operators on $L^2(\Sigma)$. Here $\dashint$ denotes the
principal value integral, as usual, and the bound
can be chosen independent of the divisor as long as it stays some finite
distance away from $\Sigma$.
\item
$C f$ is a meromorphic function off $\Sigma$, with divisor given
by $((C f)_{j1}) \ge -\dimuzs$ and $((C f)_{j2}) \ge -\dimuz$.
\item
$(Cf)(\rho^\ast) = (0\quad\ast), \qquad (Cf)(\rho) = (\ast\quad 0)$.
\end{enumerate}
Furthermore, $C$ restricts to $L^2_{s}(\Sigma)$, that is
\be
(C f) (p^\ast) = (Cf)(p) \sigI,\quad p\in\mathbb{M}\backslash\Sigma
\ee
for $f\in L^2_{s}(\Sigma)$ and if $w_\pm$ satisfy \eqref{eq:wpmsym} we also have
\be \label{eq:symcpm}
C_\pm(f w_\mp)(p^\ast) = C_\mp(f w_\pm)(p) \sigI,\quad p\in\Sigma.
\ee
\end{theorem}

Now, let the operator $C_w: L_s^2(\Sigma)\to L_s^2(\Sigma)$ be defined by
\be\label{defcw}
C_w f = C_+(f w_{-}) + C_-(f w_{+}),\quad f\in L^2_s(\Sigma),
\ee
for a $2\times2$ matrix valued $f$, where
\[
w_+ = b_+ - \id  \quad\text{ and }\quad   w_- = \id - b_-.
\]
Recall from Lemma~\ref{lem:singlesoliton} that the
unique solution corresponding to $J\equiv \id$ is given by
\[
m_0(p)= \begin{pmatrix} f(p^\ast) & f(p) \end{pmatrix}, 
\]
for some given $f(p)$ with $(f) \geq - \di_{\uhmuz} - \di_{\rho^*}$.
Since we assumed $\dimuz$ to be away from $\Sigma$, we clearly have $m_0 \in L_s^2(\Sigma)$.

\begin{theorem}[\cite{kt2,krt2}]\label{thmQ}
Assume Hypothesis~\ref{hyp:rhp} and let $m_0\in\C^2$ be given.

Suppose $m$ solves the Riemann--Hilbert problem \eqref{rhpm}. Then
\be \label{fQ}
m(p) = (1-c_0) m_0(p) + \frac{1}{2\pi\I} \int_{\Sigma} \mu(s) (w_{+}(s) + w_{-}(s)) \ul{\Omega}_{p}^{\uhmuz,\rho},
\ee
where
\[
\mu = m_+ b_+^{-1} = m_- b_-^{-1} \quad\mbox{and}\quad
c_0= \left( \frac{1}{2\pi\I} \int_{\Sigma} \mu(s) (w_{+}(s) + w_{-}(s)) \ul{\Omega}_{p_\infty}^{\uhmuz,\rho} \right)_{\!1}.
\]
Here $(m)_j$ denotes the $j$'th component of a vector.
Furthermore, $\mu$ solves
\be
(\id - C_w) \mu = (1-c_0) m_0(p).
\ee

Conversely, suppose $\ti{\mu}$ solves 
\be\label{musie}
(\id - C_w) \ti{\mu} = m_0,
\ee
and
\[
\ti{c}_0= \left( \frac{1}{2\pi\I} \int_{\Sigma} \ti{\mu}(s) (w_{+}(s) + w_{-}(s)) \ul{\Omega}_{p_\infty}^{\uhmuz,\rho} \right)_{\!1} \ne 0,
\]
then $m$ defined via \eqref{fQ}, with $(1-c_0)=(1-\ti{c}_0)^{-1}$ and $\mu=(1-\ti{c}_0)^{-1}\ti{\mu}$,
solves the Riemann--Hilbert problem \eqref{rhpm} and $\mu= m_\pm b_\pm^{-1}$.
\end{theorem}

Hence we have a formula for the solution of our Riemann--Hilbert problem $m(z)$ in terms of
$(\id - C_w)^{-1} m_0$ and this clearly raises the question of bounded
invertibility of $\id - C_w$. This follows from Fredholm theory (cf.\ e.g. \cite{zh}):

\begin{lemma}[\cite{kt2,krt2}]
Assume Hypothesis~\ref{hyp:rhp}.
Then the operator $\id-C_w$ is Fredholm of index zero,
\be
\ind(\id-C_w) =0.
\ee
\end{lemma}

By the Fredholm alternative, it follows that to show the bounded invertibility of $\id-C_w$
we only need to show that $\ker (\id-C_w) =0$. The latter being equivalent to
unique solvability of the corresponding vanishing Riemann--Hilbert problem.

\begin{corollary}\label{cor:fral}
Assume Hypothesis~\ref{hyp:rhp}.

A unique solution of the Riemann--Hilbert problem \eqref{rhpm} exists if and only if the corresponding
vanishing Riemann--Hilbert problem, where the normalization condition is given by
$m(p_{\infty})= \begin{pmatrix} 0 & 0\end{pmatrix}$, has at most one solution.
\end{corollary}

We are interested in comparing two Riemann--Hilbert problems associated with
respective jumps $w_0$ and $w$ with $\|w-w_0\|_\infty$ small,
where
\be
\|w\|_\infty= \|w_+\|_{L^\infty(\Sigma)} + \|w_-\|_{L^\infty(\Sigma)}.
\ee
For such a situation we have the following result:

\begin{theorem}[\cite{krt}]\label{thm:remcontour}
Assume that for some data $w_0^t$ the operator
\be
\id-C_{w_0^t}: L^2(\Sigma) \to L^2(\Sigma)
\ee
has a bounded inverse, where the bound is independent of $t$.

Furthermore, assume $w^t$ satisfies
\be
\|w^t - w_0^t\|_\infty \leq \alpha(t)
\ee
for some function $\alpha(t) \to 0$ as $t\to\infty$. Then
$(\id-C_{w^t})^{-1}: L^2(\Sigma)\to L^2(\Sigma)$ also exists
for sufficiently large $t$ and the associated solutions of
the Riemann--Hilbert problems \eqref{rhpm} only differ by $O(\alpha(t))$.
\end{theorem}

 \noindent
{\bf Acknowledgments.}
We want to thank Ira Egorova for most helpful discussions on this topic and pointing out several misprints
in the original version of this manuscript.

\end{document}